\documentclass[12pt,preprint]{aastex}
\begin{document}
%
%
\newcommand{\paperi}{[CD06]}
\newcommand{\dubyasqd}{W^2}
\newcommand{\tdubyasqd}{{\tilde W}^2}
\newcommand{\valfsqd}{v_{\rm A}^2}
\newcommand{\dlogrho}{\frac{d\ln\rho_0}{dz}}
\newcommand{\pgaszero}{p_0}
\newcommand{\pcosmiczero}{p_{{\rm cr,}0}}
\newcommand{\pmagzero}{\frac{B_0^2}{8\pi}}
\newcommand{\pgas}{p}
\newcommand{\tsigma}{\tilde\sigma}
\newcommand{\pcosmic}{p_{\rm cr}}
\newcommand{\rhozero}{\rho_0}
\newcommand{\pmag}{\frac{B^2}{8\pi}}
\newcommand{\gamgas}{\gamma}
\newcommand{\gamcos}{\gamma_{\rm cr}}
\newcommand{\perkpc}{\ {\rm kpc}^{-1}}
\newcommand{\permyr}{\ {\rm Myr}^{-1}}
\newcommand{\kcrit}{k_{\rm crit}}
\newcommand{\kperp}{{\bf k}_\perp}
\newcommand{\kpersqd}{k_\perp^2}
\newcommand{\kapperp}{\kappa_\perp}
\newcommand{\kappar}{\kappa_\parallel}
\newcommand{\Dpar}{D_\parallel}
\newcommand{\Dcond}{D_{\rm cond}}
\newcommand {\del}{\nabla}
\newcommand{\teta}{\tilde\eta}
\newcommand{\tnu}{\tilde\nu}
\newcommand{\omegzero}{\omega_0}
\newcommand{\omegzerosqd}{\omega_0^2}
\newcommand{\omegonesqd}{\omega_1^2}
\newcommand{\omegtwosqd}{\omega_2^2}
\newcommand{\omegthreesqd}{\omega_3^2}
\newcommand{\omegfoursqd}{\omega_4^2}
\newcommand{\omegfivesqd}{\omega_5^2}
\newcommand{\omegalfsqd}{\omega_{\rm A}^2}
\newcommand{\omegssqd}{\omega_{\rm s}^2}
\newcommand{\omegsminustwo}{\omega_{\rm s}^{-2}}
\newcommand{\omegrefsqd}{\omega_{\rm ref}^2}
\newcommand{\tomegrefsqd}{\tilde\omega_{\rm ref}^2}
\newcommand{\omegref}{\omega_{\rm ref}}
\newcommand{\tomegzerosqd}{\tilde\omega_0^2}
\newcommand{\tomegonesqd}   {\tilde\omega_1^2}
\newcommand{\tomegtwosqd}    {\tilde\omega_2^2}
\newcommand{\tomegthreesqd} {\tilde\omega_3^2}
\newcommand{\tomegfoursqd} {\tilde\omega_4^2}
\newcommand{\tomegfivesqd}    {\tilde\omega_5^2}
\newcommand{\tomegalfsqd}      {\tilde\omega_{\mathrm A}^2}
\newcommand{\tomegzero}{\tilde\omega_0}
\newcommand{\omegone}   {\omega_1}
\newcommand{\tomegone}   {\tilde\omega_1}
\newcommand{\omegtwo}    {\omega_2}
\newcommand{\tomegtwo}    {\tilde\omega_2}
\newcommand{\omegthree} {\omega_3}
\newcommand{\tomegthree} {\tilde\omega_3}
\newcommand{\omegfour} {\omega_4}
\newcommand{\tomegfour} {\tilde\omega_4}
\newcommand{\omegfive}    {\omega_5}
\newcommand{\tomegfive}    {\tilde\omega_5}
\newcommand{\omegalf}      {\omega_{\mathrm A}}
\newcommand{\tomegalf}      {\tilde\omega_{\mathrm A}}
\newcommand{\Bvec}{{\mathbf B}}
\newcommand{\kvec}{\mathbf k}
\newcommand{\vvec}{\mathbf v}
\newcommand{\rvec}{\mathbf r}
\newcommand{\gvec}{\mathbf g}
\newcommand{\bhat}{\hat {\mathbf b}}
\newcommand{\xhat}{\hat{\mathbf x}}
\newcommand{\yhat}{\hat{\mathbf y}}
\newcommand{\zhat}{\hat{\mathbf z}}
\newcommand{\specheatV}{C_V}
\newcommand{\crit}{\mathcal C}
\newcommand{\tcrit}{\tilde{\mathcal C}}
\newcommand{\routh}{{\mathcal R}}
\newcommand{\prinmin}{\mathcal M}
\newcommand{\sinsqd}{\sin^2\theta}
\newcommand{\twooverbeta}{\frac{2}{\beta}}
\newcommand{\oneoverbeta}{\frac{1}{\beta}}
\newcommand{\ptot}{p_{\rm tot}}
\newcommand{\ptotzero}{p_{\rm tot,0}}
%
%

\title{Parker/buoyancy instabilities with anisotropic thermal conduction,
cosmic rays, and arbitrary magnetic field strength}
\author{Timothy J. Dennis}
\email{ tdennis@pas.rochester.edu}
\affil{ Department of Physics \& Astronomy, 
        University of Rochester}
\author{Benjamin D. G. Chandran }
\email{ benjamin.chandran@unh.edu}
\affil{Department of Physics and Space Science Center, 
       University of New Hampshire, 
       Durham, NH}
\begin{abstract}
We report the results of a local stability analysis for a magnetized,
gravitationally stratified plasma containing cosmic rays. We account
for cosmic-ray diffusion and thermal conduction parallel to the
magnetic field and allow $\beta = 8\pi p/B^2$ to take any value, where
$p$ is the plasma pressure and $B$ is the magnetic field strength.  We
take the gravitational acceleration to be in the $-z$-direction and
the equilibrium magnetic field to be in the $y$-direction, and we
derive the dispersion relation for small-amplitude instabilities and
waves in the large-$|k_x|$ limit. We use the Routh-Hurwitz criterion
to show analytically that the necessary and sufficient criterion for
stability in this limit is $n k_B dT/dz + dp_{\rm cr}/dz + (1/8\pi)
dB^2/dz > 0$, where $T$ is the temperature, $n$ is the number density
of thermal particles, and $p_{\rm cr}$ is the cosmic-ray pressure.  We
present approximate analytical solutions for the normal modes in the
low- and high-diffusivity limits, show that they are consistent with
the derived stability criterion, and compare them to numerical results
obtained from the full, unapproximated, dispersion relation. Our
results extend earlier analyses of buoyancy instabilities in
galaxy-cluster plasmas to the $\beta\lesssim 1$~regime. Our results
also extend earlier analyses of the Parker instability to account for
anisotropic thermal conduction, and show that the interstellar medium is
more unstable to the Parker instability than was predicted by previous
studies in which the thermal plasma was treated as adiabatic.
\end{abstract}
\maketitle

\section{Introduction}
\label{sec:intro} 

Convection plays an important and well-known role in the transport of
energy in stellar interiors. It has also been argued that convection
is important in a number of low-density astrophysical plasmas, such as
the intracluster medium in clusters of galaxies (Chandran \& Rasera
2007) and accretion flows onto compact objects (Quataert \& Gruzinov
2000).  Although convection in stellar interiors has been thoroughly
studied over the course of several decades, the theory of convection
in low-density plasmas is still being developed, and investigations
carried out during the last several years have led to some interesting
surprises.

For many years, it was widely assumed that the convective stability
criterion for a low-density, non-rotating, weakly magnetized plasma is
the Schwarzchild criterion, $ds/dz>0$, where $s$ is the specific
entropy of the plasma and the gravitational acceleration is in
the~$-z$ direction.  However, Balbus~(2000, 2001) showed that even
weak magnetic fields strongly modify the convective stability
criterion by causing heat to be conducted almost exclusively along
magnetic field lines.  This anisotropy in the thermal conductivity
arises when the electron gyroradius is much less than the electron
mean free path, a condition that is easily satisfied for realistic
magnetic fields in most cases of interest.  Balbus considered an
equilibrium in which the magnetic field is in the~$xy$-plane, and in
which $\beta = 8\pi p/B^2 \gg 1$, where $p$ is the pressure and~$B$ is
the magnetic field strength.  He showed that near marginal stability,
the temperature of a rising fluid parcel is almost constant,
essentially for two reasons. First, the parcel remains magnetically
connected to material at its initial height. Second, near marginal
stability a fluid parcel rises very slowly, so thermal conduction has
enough time to approximately equalize the temperature along the
perturbed magnetic field lines. As a result, the stability criterion
becomes~$dT/dz>0$.  When this criterion is satisfied, a rising fluid
parcel is cooler than its surroundings, and hence denser at the same
pressure, so that it falls back down to its initial height. 
Parrish \& Stone (2005, 2007) carried out numerical
simulations that validated Balbus' analysis and extended it to
the nonlinear regime.

An immediate question arises, namely, why doesn't Balbus's stability
criterion apply to stars?
Although stellar plasmas are magnetized, heat is conducted in
stellar interiors primarily by photons. As discussed by Balbus (2000,
2001), the conductivity is thus almost isotropic, so it is the
Schwarzchild criterion that applies. The reason that the Schwarzchild
criterion applies even if the isotropic conductivity is large is
somewhat subtle. If $ds/dz<0$, then adiabatic expansion would cause a
slowly rising fluid parcel to be hotter and lighter than its
surroundings, and hence buoyant.  The effect of isotropic conductivity
is then to relax the temperature in the parcel towards that of the
immediately surrounding fluid. However, because the conductivity is
finite, the rising fluid parcel's temperature is never decreased all
the way to the temperature of its surroundings.  The fluid parcel thus
remains slightly hotter than its surroundings, and hence slightly
lighter at the same pressure, and the fluid is convectively unstable.
Although isotropic conductivity does not modify the Schwarzchild
stability criterion, it does reduce the convective heat flux and the
``efficiency of convection'' in a convectively unstable fluid by
decreasing the temperature difference $\delta T$ between rising fluid
parcels and their surroundings (Cox \& Guili 1968).

Balbus's analysis has been extended in two ways by recent
studies. First, Chandran \& Dennis (2006), (hereafter referred to as 
\paperi), investigated how the stability criterion is affected by 
the presence of cosmic rays that diffuse primarily along magnetic 
field lines. Like Balbus (2000, 2001), they assumed that~$\beta \gg 1$
and took the equilibrium magnetic field to be in the~$xy$-plane They 
showed that near marginal stability, the cosmic-ray pressure is nearly 
constant within a rising fluid element. This is because the fluid element
remains connected to material at its initial height, and because fluid 
elements rise very slowly near marginal stability, so that there is plenty
of time for cosmic-ray diffusion to approximately equalize the cosmic-ray
pressure~$p_{\rm cr}$ along the perturbed magnetic field lines.
\paperi\ showed analytically that the stability criterion in the presence
of cosmic rays is $n k_B dT/dz + dp_{\rm cr}/dz > 0$, where $n$ is the
total number density of thermal particles. 

More recently, Quataert (2007) considered buoyancy instabilities in a
low-density, high-$\beta$ plasma in the absence of cosmic rays, but
allowing the equilibrium magnetic field to have a component in the~$z$
direction, parallel or antiparallel to the direction of gravity. Since
the temperature is a function of~$z$, the $z$ component of the
equilibrium magnetic field leads to an equilibrium heat flux.
Quataert (2007) showed that this heat flux causes the plasma to become
convectively unstable even if~$dT/dz>0$, so that the plasma is always
convectively unstable if the magnetic field has a nonzero $z$
component and a nonzero component in the $xy$ plane.  This
heat-flux-buoyancy instability arises because of the geometry of the
perturbed magnetic field lines in the plasma. For example, when the
magnetic field is in the~$z$ direction and a fluid element is
displaced upwards at a 45-degree angle with respect to the $z$ axis,
field lines converge as they enter the fluid element from ``above''
(i.e., from larger~$z$).  As a result, if $dT/dz>0$ then the parallel
heat flux converges within the fluid element, causing the fluid
element to become hotter than its surroundings, and thus less dense at
the same pressure. Buoyancy forces then cause the upwardly displaced
fluid element to rise unstably. (Quataert~2007) The nonlinear
development of this instability was investigated numerically by
Parrish \& Quataert (2007).

One of the open questions in this area of research is whether the
buoyancy instabilities identified in these previous studies for
the~$\beta \gg 1$ regime still operate when the magnetic field
strength is increased to the point that~$\beta \lesssim 1$. We address
this question in this paper. We consider the equilibrium geometry
investigated by Balbus (2000, 2001) and \paperi, in
which the magnetic field is in the~$xy$-plane - in particular, we set
$\Bvec_0 = B_0 \yhat$. We also allow for cosmic rays that diffuse
along magnetic field lines, but now we allow~$\beta$ to take any
value.  We focus on wave vectors in the ``quasi-interchange'' limit,
in which $|k_x| \gg |k_y|$, $|k_x| \gg |k_z|$, and $|k_x H| \gg 1$,
where~$H$ is the density scale height. This is the most unstable
wave-vector regime for stratified adiabatic plasmas, because a
small~$k_y$ reduces the stabilizing effects of magnetic tension and a
large~$k_x$ allows a rising fluid element to easily get out of the way
of the next rising element beneath it by moving just a small distance
in the~$x$ direction. (Parker 1967, Shu 1974, Ferri\`ere et al. 1999)
We show analytically that the stability criterion in this limit is
\begin{equation}
nk_B \frac{dT}{dz} + 
\frac{d p_{\rm cr}}{d z}
+ \frac{1}{8\pi} \frac{dB^2}{dz} > 0,
\label{eq:stabcrit0}
\end{equation}
and we present a heuristic derivation of this stability criterion from
physical arguments.  We also derive approximate analytical solutions
to the dispersion relation for small-amplitude perturbations to the
equilibrium in different parameter regimes, and compare these
solutions to numerical solutions of the full dispersion relation.

Our results are important for determining the convective stability of
galaxy-cluster plasmas, in which cosmic-rays are often produced by
central radio sources.  Convection in intracluster plasmas is of
interest because it may provide a mechanism for regulating the
temperature profiles of galaxy-cluster plasmas and offsetting
radiative cooling, thereby solving the so-called ``cooling-flow
problem.''  (Chandran 2004, 2005; Parrish \& Stone 2005, 2007;
Chandran \& Rasera 2007). Our results are also important for
determining the conditions under which the Parker instability can
operate in the interstellar medium (ISM). Previous treatments of the
Parker instability assume an adiabatic thermal plasma (Parker 1966,
1967, Shu 1974, Ryu 2003).  Our results show that anisotropic thermal
conductivity makes the ISM more unstable to the Parker instability, so
that the instability can operate under a wider range of equilibrium
profiles than was previously recognized.

The remainder of this paper is organized as follows. In section
\ref{sec:disprel} we outline the derivation of the general form of the
dispersion relation.  In section \ref{sec:qil} we specialize to the
quasi-interchange limit, present our derivation of the necessary and
sufficient condition for convective stability, and describe the
properties of the unstable eigenmodes in plasmas that are
very close to marginal stability. In section~\ref{sec:physical} we
present a heuristic, physical derivation of the stability criterion.
We discuss the implications of our work for galaxy-cluster plasmas and
the interstellar medium in sections~\ref{sec:gc} and~\ref{sec:parker},
respectively.  In section~\ref{sec:conc} we summarize our results, and
in appendix~\ref{ap:eigenmodes} we present approximate analytic
solutions and numerical solutions to the dispersion relation.

\section{The general dispersion relation}
\label{sec:disprel}
We begin with a standard set of two-fluid equations 
(Drury \& Volk 1981, Jones~\&~Kang 1990), which we modify 
to include thermal conduction along the magnetic field: 
\begin{eqnarray}
\label{eq:masscon}
\frac{d\rho}{dt}&=&-\rho\del\cdot\vvec,      \\
\label{eq:momconv}
\frac{d\vvec}{dt}&=&-\frac{1}{\rho}\del
\left(
 \pgas+\pcosmic+\pmag
\right) +
 \frac{1}{4\pi\rho}\Bvec\cdot\del\Bvec 
+ \gvec,                                     \\
\label{eq:induc}
 \frac{d\Bvec}{dt}&=&-\Bvec\del\cdot\vvec 
+ \Bvec\cdot\del\vvec,                       \\
\label{eq:pgas}
\frac{d\pgas}{dt}&=&-
\gamgas\pgas\del\cdot\vvec +
\left(
 \gamgas-1
\right)\del\cdot
\left[
 \bhat\kappar
 \left(
  \bhat\cdot\del T
 \right)
\right],                                     \\
\label{eq:pcosmic}
\frac{d\pcosmic}{dt}&=&
-\gamcos\pcosmic\del\cdot\vvec+\del\cdot
\left[
 \bhat\Dpar
 \left(
  \bhat\cdot\del\pcosmic
 \right)
\right],
\end{eqnarray}
where $d/dt=\partial/\partial t + \vvec\cdot\nabla$, and 
where $\rho$ is the plasma mass density, $\vvec$ is the 
velocity, $\pgas$ is the plasma pressure, $\pcosmic$ is 
the cosmic-ray pressure,  $\gamgas$ is the ratio of specific 
heats of the plasma, $\gamcos$ is the effective ratio of 
specific heats for the cosmic rays, $\Bvec$ is the magnetic 
field, $\bhat$ is a unit vector in the direction of the 
magnetic field, $\kappar$ is the thermal conductivity along 
the direction of the magnetic field, $\Dpar$ is the 
cosmic-ray diffusivity along the direction of the magnetic 
field, and $\gvec$ is the gravitational acceleration. We have 
ignored cross-field conduction and diffusion in 
equations~(\ref{eq:pgas})~and~(\ref{eq:pcosmic}) since the gyroradii 
of the thermal particles are small compared to their Coulomb 
mean free path, and the gyroradii of the cosmic rays are 
small compared to the mean free path for cosmic-ray 
scattering. Equations~(\ref{eq:masscon})$-$(\ref{eq:pcosmic})
are closed via the equation of state for an ideal gas:
\begin{equation}
\label{eq:eqstate}
\pgas=\specheatV
\left(
 \gamgas-1
\right)\rho T.
\end{equation}
We take
\begin{equation}
\label{eq:equilg}
\gvec=-g\zhat,
\end{equation}
and consider an equilibrium in which
\begin{equation}
\label{eq:equilB}
\Bvec_0=B_0\yhat.
\end{equation}
All equilibrium quantities (denoted with a ``0'' subscript)
are taken to be functions of $z$
only, and we set $\vvec_0=0$. These assumptions lead to a 
condition for hydrostatic equilibrium of the form:
\begin{equation}
\label{eq:hydroeq}
\frac{d}{dz}
\left(
 \pgaszero+\pcosmiczero+\pmagzero
\right)
=-\rhozero g.
\end{equation}
To introduce perturbations, we represent the variables in 
our two-fluid equations as sums of an equilibrium value and 
a small fluctuating quantity as follows:
\begin{eqnarray}
\label{eq:rhoperturb}
\rho &=& \rhozero+\delta\rho                 \\
\label{eq:pgasperturb}
\pgas &=& \pgaszero+\delta\pgas,             \\
\nonumber\cdots
\end{eqnarray}
We employ a local analysis, in which we take the fluctuating
quantities to be proportional to
$e^{i\left(\kvec\cdot\rvec-\omega t\right)}$, with 
\begin{equation}
kH \gg 1,
\label{eq:local1} 
\end{equation} 
where
\begin{equation}
H \equiv \left|\frac{d \ln \rho_0}{dz}\right|^{-1}
\label{eq:defH} 
\end{equation} 
is the density scale height, which we take to be comparable
to the length scales over which each of the equilibrium quantities
varies. Substituting 
equations (\ref{eq:rhoperturb}) and (\ref{eq:pgasperturb}), 
and analogous expressions for 
$\pcosmic$, $\Bvec$, $\vvec$, $\bhat$, and $T$ into 
equations~(\ref{eq:masscon})$-$(\ref{eq:pcosmic}), and into 
equation~(\ref{eq:eqstate}), we obtain the following 
equations for the fluctuating quantities:
\begin{equation}
\label{eq:linmasscon}
-i\omega\frac{\delta\rho}{\rho_0}
+i\kvec\cdot\delta\vvec
+\delta v_z\dlogrho=0,
\end{equation}
\begin{equation}
\label{eq:linmomcon}
-i\omega\delta\vvec=-g\frac{\delta\rho}{\rho_0}\zhat 
-i\kvec\frac{\delta\ptot}{\rho_0}
+\valfsqd
\left[
 \frac{d\ln B_0}{dz}\frac{\delta B_z}{B_0}\yhat
 +ik_y\frac{\delta\Bvec}{B_0}
\right],
\end{equation}
\begin{equation}
\label{eq:lininduc}
-i\omega\frac{\delta\Bvec}{B_0}
=ik_y\delta\vvec-\delta v_z\frac{d\ln B_0}{dz}
-\yhat\left(i\kvec\cdot\delta\vvec\right),
\end{equation}
\begin{equation}
\label{eq:linpgas}
-i\omega
\left[
 \frac{\delta\pgas}{\pgaszero}
-\gamgas\frac{\delta\rho}{\rho_0}
\right]+ 
\delta v_z \frac{d\ln
\left(
 \pgas\rho^{-\gamgas}
\right)}{dz} = \Dcond
\left[ 
 ik_y\frac{d\ln T_0}{dz}\frac{\delta B_z}{B_0}
- k_y^2\frac{\delta T}{T_0}
\right],
\end{equation}
\begin{equation}
\label{eq:linpcosmic}
-i\omega\frac{\delta\pcosmic}{\pcosmiczero} 
+ \delta v_z\frac{d\ln\pcosmiczero}{dz} 
+ i\gamcos\kvec\cdot\delta\vvec =\Dpar
\left[
 ik_y\frac{d\ln\pcosmiczero}{dz}
      \frac{\delta B_z}{B_0} 
 - k_y^2\frac{\delta\pcosmic}{\pcosmiczero}
\right],
\end{equation}
\begin{equation}
\label{eq:lineqstate}
\frac{\delta\pgas}{\pgaszero}
=\frac{\delta\rho}{\rho_0}
+\frac{\delta T}{T_0},
\end{equation}
where,
\begin{equation}
\label{eq:Dcond}
\Dcond=\frac{\left(\gamgas-1\right)\kappar T_0}{\pgaszero},
\end{equation}
and 
\begin{equation}
\label{eq:ptotdef}
\ptot=\pgas+\pcosmic+\frac{B^2}{8\pi}.
\end{equation}
Equations (\ref{eq:linmasscon})--(\ref{eq:lineqstate})
may be reduced to an expression of the form:
${\mathsf M}\cdot\delta\vvec=0,$
where $\mathsf M$ is a $3\times3$ matrix. For non-trivial 
solutions of this equation, we require 
$|{\sf M}|=0$, whence we obtain the dispersion relation:
\begin{equation}
\label{eq:disprel}
A_0\omega^6+A_2\omega^4+A_4\omega^2+A_6=0,
\end{equation}
where,
\begin{eqnarray}
\label{eq:bigA0}
A_0 &=& 1,                                    \\
\label{eq:bigA2}
A_2&=& -k^2
\left(
 u^2+\valfsqd
\right)
-k_y^2\valfsqd+g\dlogrho,                    \\
\label{eq:bigA4}
A_4&=& k_y^2k^2\valfsqd
\left(
 2u^2+\valfsqd
\right)-
\left(
 k_x^2+k_y^2
\right)
\left[
g^2+
\left(
u^2+\valfsqd
\right) g\dlogrho
\right],                                    \\
\label{eq:bigA6}
A_6&=&k_y^2\valfsqd
\left[
-k^2k_y^2\valfsqd u^2 + 
\left(
 k_x^2+k_y^2
\right)
\left(
 g^2+u^2g\dlogrho
\right)
\right],
\end{eqnarray}
and
\begin{equation}
\label{eq:udef}
u^2=\frac{1}{\rho_0}
\left[
\pgaszero
\left(
 \frac{\gamgas\omega+i\eta}{\omega+i\eta}
\right)+
\pcosmic\frac{\gamcos\omega}{\omega+i\nu}
\right],
\end{equation}
and where in equation~(\ref{eq:udef}) we have introduced the 
quantities
\begin{eqnarray}
\eta&=&k_y^2\Dcond,                          \\
 \nu&=&k_y^2\Dpar,                           \\
\valfsqd &=&\frac{B_0^2}{8\pi\rho_0},
\end{eqnarray}
where $v_{\rm A}$ is the Alfv\`en speed, and $\eta$ and $\nu$ are,
respectively, the rates at which temperature fluctuations and
cosmic-ray-pressure fluctuations are smoothed out along the magnetic
field.  Equations (\ref{eq:disprel})--(\ref{eq:udef}) represent the
same result as that presented in equations~(26)~and~(27) of
\paperi\ and we shall henceforth refer to this result as the ``general
dispersion relation.''  As we shall see, this relation constitutes an
eighth-order polynomial equation in $\sigma=-i\omega$, (where the
change of variables is made so as to make all of the polynomial
coefficients real). It is worthwhile to note that in the absence of
cosmic rays and thermal conductivity,
\begin{equation}
\label{eq:limu}
u^2
\stackrel{\displaystyle\longrightarrow}
{\scriptstyle \nu,\eta,\pcosmic\rightarrow 0}
{}\frac{\gamgas\pgaszero}{\rho_0}=c_s^2,
\end{equation}
where $c_s$ is the adiabatic sound speed, and that if we 
take this limit, together with the limit of no 
stratification and $g\rightarrow 0$, the general dispersion relation 
reduces to the well-known dispersion relation obtained in 
ideal MHD. Thus the normal modes described by equation~(\ref{eq:disprel}) 
may be viewed as modifications of the 
Alfv\`en mode and the  fast and slow magnetosonic modes
of ideal MHD. 

We now present the definitions of a number of frequencies 
that allow us to write the polynomial form of the dispersion
relation more compactly. These are:
\begin{eqnarray}
\omegssqd&=&\frac{k^2\pgaszero}{\rho_0},     
\label{eq:omegassqd} \\
\omegalfsqd&=&k_y^2\valfsqd,                 \\
\omegzerosqd&=&\frac{\rho_0}{\pgaszero}g^2
\sin^2\theta,                                \\
\omegonesqd&=&\frac{g\sin^2\theta}{\gamgas} 
\frac{d}{dz}\ln
\left(\pgaszero\rho_0^{-\gamgas}
\right),                                     \\
\omegtwosqd&=&g\sin^2\theta\frac{d}{dz}\ln T,\label{eq:omegatwosqd} \\
\omegthreesqd&=&\frac{g\sin^2\theta}{\gamcos} 
\frac{d}{dz}\ln
\left(\pcosmiczero\rho_0^{-\gamcos}
\right),                                     \\
\omegfoursqd&=&
g\sin^2\theta\frac{d}{dz}\ln\pcosmiczero, \label{eq:omegafoursqd} \hspace{0.3cm} \mbox{ and}\\
\omegfivesqd&=&
g\sin^2\theta\frac{d}{dz}\ln B_0^2,
\label{eq:omegafivesqd} \\
\end{eqnarray}
where we have defined 
\begin{equation}
\label{eq:sinsqd}
\sin^2\theta=\frac{k_x^2+k_y^2}{k^2}.
\end{equation}
The quantities $\omegalfsqd$ and $\omegssqd$ are the 
squares of the Alfv\`en and isothermal sound-wave 
frequencies respectively. The quantity $\omegonesqd$ is 
the square of the usual Brunt-V\"ais\"al\"a  frequency for 
buoyancy oscillations in the limit of vanishing 
cosmic-ray pressure and magnetic field. As we shall 
see below, the quantities $\omegthreesqd$ and 
$\omegfivesqd$ serve to modify the frequency of these 
oscillations when the cosmic-ray pressure and magnetic 
field are non-vanishing. The quantities 
$\omegtwosqd$ and $\omegfoursqd$ are related to $\omegonesqd
$ and $\omegthreesqd$ through the identities: 
\begin{eqnarray}
\label{eq:ident12}
\omegtwosqd&=&\gamgas\omegonesqd+
\left(
 \gamgas-1
\right)g\sinsqd\dlogrho,                      \\
\label{eq:ident34}
\omegfoursqd&=&\gamcos\omegthreesqd
+\gamcos g\sinsqd\dlogrho.
\end{eqnarray}
We also define the quantities $\dubyasqd$ and $\crit$:
\begin{equation}
\label{eq:dubyadef}
\dubyasqd=\omegtwosqd
+\chi\omegfoursqd
+\frac{1}{\beta}\omegfivesqd,
\end{equation}
and, 
\begin{equation}
\label{eq:defcrit}
\crit=\omegalfsqd+ \dubyasqd,
\end{equation} 
where in equation (\ref{eq:dubyadef}), $\chi=\pcosmiczero/\pgaszero$.
Finally, noting the identity
\begin{equation}
\label{eq:critident}
-g\sinsqd\dlogrho=\omegzerosqd
-\omegalfsqd+\crit,
\end{equation}
we find that
the general dispersion relation may be written
\begin{equation}
\label{eq:gendisp}
a_0\sigma^8+a_1\sigma^7+\cdots+a_7\sigma+a_8=0,
\end{equation}
where
\begin{eqnarray}
\label{eq:smallazero}
a_0&=&\omegsminustwo,                        \\
\label{eq:smallaone}
a_1&=&
\left(
\nu+\eta
\right)\omegsminustwo,                        \\
\label{eq:smallatwo}
a_2&=&
\left(
 \gamgas+\chi\gamcos+\frac{2}{\beta}
\right)+
\left[
 \nu\eta+\omegalfsqd-g\dlogrho
\right] \omegsminustwo                       \\
\label{eq:smallathree}
a_3&=&
\left[
 \nu
\left(
 \gamgas+\frac{2}{\beta}
\right)
+\eta
\left(
1+\chi\gamcos+\frac{2}{\beta}
\right)
\right] +
\left[
\left(
 \nu+\eta
\right)
\left(
\omegalfsqd-g\dlogrho
\right)
\right]
\omegsminustwo,                              \\
\label{eq:smallafour}
a_4&=& \nu\eta
\left[
\left(
 1+\frac{2}{\beta}
\right)+
\left(
\omegalfsqd-g\dlogrho
\right)\omegsminustwo
\right] + \nonumber                            \\
&{}&\quad
\left[
  \left(
    \left[
      \left(
        \gamgas-1
      \right)+\chi\gamcos
    \right]+\frac{2}{\beta}
  \right)\omegzerosqd+
  \left(
    \gamgas+\chi\gamcos
  \right)\omegalfsqd+
  \left(
    \gamgas+\chi\gamcos+\frac{2}{\beta}
  \right)\crit
\right],                                     \\
\label{eq:smallafive}
a_5&=& \nu
\left[
\left(
 \gamgas+\frac{2}{\beta}
\right)
\left(
 \omegzerosqd+\crit
\right)-\omegzerosqd+\gamgas\omegalfsqd
\right]+ \nonumber                             \\
&{}&\qquad\qquad\eta 
\left[
\left(
 1+\chi\gamcos+\twooverbeta
\right)
\left(
 \omegzerosqd+\crit
\right)-\omegzerosqd +
\left(
 1+\chi\gamcos
\right)\omegalfsqd
\right],                                     \\
\label{eq:smallasix}
a_6&=&\nu\eta
\left[
\left(
 1+\twooverbeta
\right)
\left(
 \omegzerosqd+\crit
\right)-
\left(
 \omegzerosqd-\omegalfsqd
\right)
\right]  + \omegalfsqd
\left[
\left(
 \gamgas+\chi\gamcos
\right)
\left(
 \omegzerosqd+\crit
\right)-\omegzerosqd
\right],                                     \\
\label{eq:smallaseven}
a_7&=& \omegalfsqd
\biggl(
\left[
 \nu
 \left(
   \gamgas-1
 \right)+\chi\gamcos\eta
\right]\omegzerosqd+
\left[
 \nu\gamgas+\eta
  \left(
    1+\chi\gamcos
  \right)
\right]\crit
\biggr),                             \hspace{0.3cm} \mbox{ and}        \\
\label{eq:smallaeight}
a_8&=&\nu\eta\omegalfsqd\crit.
\end{eqnarray}

We assume that $|d\ln\pgaszero/dz|$, $|d\ln\pcosmiczero/dz|$, and
$|d\ln B_0^2/dz|$ are of order $H^{-1}$. We may thus conclude from
equation~(\ref{eq:hydroeq}) that $g\sim\ptotzero H^{-1}/\rho_0$. We
also assume that $p_{\rm tot}$ is not much greater than~$p$. Our
assumption that $|kH \gg 1|$ then allows us to write that
\begin{equation}
\label{eq:dropterm}
\omegsminustwo g\dlogrho\sim
\left(
 k^2H^2
\right)^{-1}\ll1. 
\end{equation}
This inequality enables us to drop the $\omega_s^{-2}g\,d\ln\rho_0/dz$
terms in equations (\ref{eq:smallatwo}), (\ref{eq:smallathree}), and
(\ref{eq:smallafour}). 

As a check on the results of this section, we show in appendix
\ref{ap:pslimit} that equation (\ref{eq:gendisp}) reduces properly to
the results obtained by Parker (1966, 1967) and Shu (1974) when
cosmic-ray diffusivity is taken to be infinite and thermal conduction
is negligible, and when the results of Parker (1966, 1967) and Shu (1974)
are considered in the short-wavelength limit.

\section{The quasi-interchange limit}
\label{sec:qil}

The most unstable modes in a gravitationally 
stratified adiabatic plasma threaded by a horizontal magnetic 
field are those for which $|k_x|$ is very large, so that
\begin{eqnarray}
|k_x H| & \gg & 1,\\
|k_x| & \gg & |k_y|, \\
|k_x| & \gg & |k_z|, \mbox{ \hspace{0.3cm} and}\\
\sin^2\theta & \rightarrow & 1 
\end{eqnarray} 
(Parker 1967, Shu 1974, Ferri\`ere et al. 1999).
We conjecture that the same is true when thermal conduction
is taken into account, at least when the equilibrium magnetic
field is horizontal, and thus we focus on this limit, which
we call the ``quasi-interchange limit.''

For very large~$|k_x|$, one set of modes consists of high-frequency
magnetosonic-like waves. In the $\beta \gg 1$~limit, these waves are
stable \paperi, and we assume they are stable here
as well. [We note, however, that in the presence of an equilibrium
  heat flux (i.e. $B_{0z} \neq 0$), anisotropic conduction can cause
  magnetosonic waves to become overstable (Socrates, Parrish, \& Stone
  2007).]  To filter out these high-frequency waves, we assume that
\begin{equation}
\sigma \ll \omega_s.
\label{eq:balassumps1} 
\end{equation} 
We also assume that $|k_x/k_y|$ is sufficiently large that
\begin{eqnarray}
\frac{\omega_{\rm A}}{\omega_{\rm s}} 
&\ll& 1,                                     
\label{eq:balassumps2} \mbox{ \hspace{0.3cm} and}\\
     \frac{\nu}{\omega_{\rm s}} 
\sim \frac{\eta}{\omega_{\rm s}}&\ll&1,      
\label{eq:balassumps3} 
\end{eqnarray}
and that $|k_x H|$ is sufficiently large that 
\begin{eqnarray}
\frac{\omega_i}{\omega_{\rm s}} &\ll & 
1;\hskip 1cm i = 0,\ldots, 5.
\label{eq:balassumps4} 
\end{eqnarray}
Using these inequalities and equation~(\ref{eq:dropterm}),
we can rewrite the general dispersion relation as a 
6th-degree polynomial equation,
\begin{equation}
\label{eq:ballim}
  b_0\sigma^6 
+ b_1\sigma^5 
+ b_2\sigma^4 
+ b_3 \sigma^3 
+ b_4\sigma^2
+ b_5\sigma
+ b_6=0,
\end{equation}
where
\begin{eqnarray}
\label{eq:balloonb0}
b_0&=&\gamgas+\chi\gamcos+\twooverbeta,      \\
\label{eq:balloonb1}
b_1&=& \nu
\left(
 \gamgas+\twooverbeta
\right)+\eta
\left(
 1+\chi\gamcos+\twooverbeta
\right),                                     \\
\label{eq:balloonb2}
b_2&=&
\left(
 \gamgas+\chi\gamcos
\right)
\left(
 \omegalfsqd+\omegzerosqd+\crit
\right)
-\omegzerosqd+\twooverbeta
\left(
 \omegzerosqd+\crit
\right)+\nu\eta
\left(
1+\twooverbeta
\right),                                     \\
\label{eq: balloonb3} 
b_3&=&
\left[
 \eta
 \left(
  1+\chi\gamcos
 \right)
+\nu\gamgas
\right]
\left(
\omegalfsqd+\omegzerosqd+\crit
\right)
+
\left(
 \nu+\eta
\right)
\left[
 \twooverbeta
\left(
 \omegzerosqd+\crit
\right)
-\omegzerosqd
\right],\qquad                               \\
\label{eq:balloonb4}
b_4&=&\omegalfsqd
\biggl[
\left(
 \gamgas+\chi\gamcos
\right)
\left(
 \omegzerosqd+\crit
\right)-\omegzerosqd
\biggr]
+\nu\eta
\left[
 \omegalfsqd+\crit+\twooverbeta
 \left(
  \omegzerosqd+\crit
 \right)
\right],                                     \\
\label{eq:balloonb5}
b_5&=&
\omegalfsqd
\biggl[
\left(
 \omegzerosqd+\crit
\right)
\left[
\nu\gamgas+\eta
\left(
 1+\chi\gamcos
\right)
\right]
-\left(
 \nu+\eta
\right)
\omegzerosqd
\biggr],                                  \hspace{0.3cm} \mbox{ and} \\
\label{eq:balloonb6}
b_6&=&\nu\eta\omegalfsqd\crit. 
\end{eqnarray}

\subsection{Stability 
            criterion}
\label{sec:stabcrit}

To obtain the stability criterion for the modes
described by equation~(\ref{eq:ballim}), we use
the Routh-Hurwitz theorem.
[see for example  Levinson \& Redheffer (1970)]
To apply this theorem, we construct the
matrix~$\routh$ from the (real) coefficients of the 
polynomial in 
equation~(\ref{eq:ballim}), where
\begin{equation}
\routh=\left(
\begin{array}{cccccc}
b_1 & b_3 & b_5 & 0   & 0   & 0   \cr
b_0 & b_2 & b_4 & b_6 & 0   & 0   \cr
  0 & b_1 & b_3 & b_5 & 0   & 0   \cr
  0 & b_0 & b_2 & b_4 & b_6 & 0   \cr
  0 &   0 & b_1 & b_3 & b_5 & 0   \cr
  0 &   0 & b_0 & b_2 & b_4 & b_6 
\end{array}
\right).
\end{equation}
The Routh-Hurwitz theorem then states that for the real parts of the
roots of equation~(\ref{eq:ballim}) to all take on negative values, it
is a necessary and sufficient condition that the determinants of the
principle minor matrices $\prinmin_i$ of $\routh$ all be
positive-definite.  This necessary and sufficient condition is the
stability criterion for our plasma.  The determinants of the principle
minors of $\routh$ are:
\begin{equation}
\det(1)=b_1,
\end{equation}
\begin{equation}
\det(2)=\left|
\begin{array}{cc}
b_1 & b_3 \cr
b_0 & b_2
\end{array}
\right|,
\end{equation}
\begin{equation}
\det(3)=\left|
\begin{array}{ccc}
b_1 & b_3 & b_5 \cr
b_0 & b_2 & b_4 \cr
  0 & b_1 & b_3 
\end{array}
\right|,
\end{equation}
\begin{equation}
\det(4)=\left|
\begin{array}{cccc}
b_1 & b_3 & b_5 &   0 \cr
b_0 & b_2 & b_4 & b_6 \cr
  0 & b_1 & b_3 & b_5 \cr
  0 & b_0 & b_2 & b_4
\end{array}
\right|,
\end{equation}
\begin{equation}
\det(5)=\left|
\begin{array}{ccccc}
b_1 & b_3 & b_5 & 0   & 0   \cr
b_0 & b_2 & b_4 & b_6 & 0   \cr
  0 & b_1 & b_3 & b_5 & 0   \cr
  0 & b_0 & b_2 & b_4 & b_6 \cr
  0 &   0 & b_1 & b_3 & b_5 
\end{array}
\right|,
\end{equation}
and
\begin{equation}
\det(6)=\left|\routh\right|.
\end{equation}
After some algebra, we find that these determinants may be 
expressed as
\begin{eqnarray}
\label{eq:rhdet1}
\det(1)&=&\nu
\left(
 \gamgas+\twooverbeta
\right)+\eta
\left(
 1+\chi\gamcos+\twooverbeta
\right),
\label{eq:rhdet2} \\
\det(2)&=&
b_1\nu\eta
\left(
 1+\twooverbeta
\right)+J
\left(
 \omegzerosqd+\twooverbeta\omegalfsqd
\right), \\
\label{eq:rhdet3} 
\det(3)&=&
b_1J\omegzerosqd
\left(
 \omegalfsqd+\dubyasqd
\right)+JK
\left(
 \omega_0^4+\twooverbeta\omega_{\rm A}^4
\right)
+\nu\eta Kb_1
\left(
 \omegzerosqd+\twooverbeta\omegalfsqd
\right) \nonumber                            \\
&{}&\qquad\qquad\qquad+\twooverbeta J
\left(
 \nu+\eta
\right)
\left(
 \omegzerosqd-\omegalfsqd
\right)^2 \\
\label{eq:rhdet4}
\det(4)&=&\twooverbeta 
J^2\omegalfsqd\omegzerosqd
\left[ 
 \dubyasqd+\omegzerosqd
\right]^2+ \nonumber                         \\
&{}&J\nu\eta
\Biggl\{
(\nu+\eta)\omega_0^2\left[\crit + \twooverbeta(\dubyasqd+\omega_0^2)\right]^2
+ K(\crit^2\omega_0^2 + \crit \omega_0^4)
 + \nonumber                         \\
&{}& \twooverbeta(\nu+\eta)\omega_A^2(\omega_A^2-\omega_0^2)^2
+ \twooverbeta K\left[ (\dubyasqd + \omega_0^2)^2\omega_0^2 + \crit \omega_0^2\omega_A^2 + \omega_A^2 (\omega_A^2-\omega_0^2)^2\right]
\Biggr\} \nonumber                           \\
&{}&+
\left(
 \nu\eta
\right)^2b_1K
\left[
\left(
1+\twooverbeta
\right)\omegzerosqd\crit+\twooverbeta
\left(
 \omegzerosqd-\omegalfsqd
\right)^2
\right], \\
\label{eq:rhdet5}
\det(5)&=&
\twooverbeta\omegzerosqd\omegalfsqd
\left[
 \dubyasqd+\omegzerosqd
\right]^2 \times\nonumber                    \\
&{}&
\biggl\{
\left(
 \nu\eta
\right)^2
b_1K^2+
\left(
 \nu\eta
\right)JK
\left[
b_1\crit+\omegalfsqd
\left(
 \nu+\eta
\right)+k
\left(
 \omegzerosqd+\omegalfsqd
\right)+\twooverbeta\omegzerosqd
\left(
 \nu+\eta
\right)
\right]\nonumber                             \\
&{}&\qquad+J^2\omegalfsqd
\left[
K\omegzerosqd+
\left( 
 K+\nu+\eta
\right)\crit
\right]
\biggr\} \hspace{0.3cm} \mbox{ and}\\
\det(6)&=&
\left|
 {\mathcal R}
\right|=b_6\det
\left(
 5
\right). \label{eq:det6eq} 
\end{eqnarray}
The quantities $J$ and $K$ appearing in the above 
expressions are defined as
\begin{equation}
\label{eq:defJ}
  J =\left(\gamgas-1\right)\eta+\chi\gamcos\nu,
\end{equation}
and
\begin{equation}
\label{eq:defK}
  K =\left(\gamgas-1\right)\nu+\chi\gamcos\eta,
\end{equation}
and are always positive.

We first consider the case $k_y \neq 0$.
In this case, $J$, $K$, and the first two
determinants are seen to be composed of sums of 
positive-definite quantities and so are themselves 
positive definite. By inspection, 
$\det(3)$  through $\det(6)$ are positive if $\crit>0$, and
thus $\crit>0$ is a sufficient condition for stability.
On the other hand, equation~(\ref{eq:det6eq}) shows that
if $\crit<0$, then either $\det(5)$ or $\det(6)$ is negative.
Therefore, $\crit>0$ is also a necessary condition for stability.
If we fix the wavevector~$\kvec$, taking~$k_y \neq 0$,
the necessary and sufficient condition for modes at that~$\kvec$ to
be stable is then
\begin{equation}
\crit >0.
\label{eq:stabcritky} 
\end{equation} 
Since $\crit = \dubyasqd + k_y^2 v_A^2$,
the smallest value of~$\crit$ is obtained in the limit~$k_y\rightarrow 0$.
The necessary and sufficient condition for the plasma
to be stable at all wavevectors in the quasi-interchange limit is thus
\begin{equation}
\dubyasqd > 0.
\label{eq:stabcritgen} 
\end{equation} 
Using the definition of $\dubyasqd$ given in 
equation~(\ref{eq:dubyadef}), and the definitions 
of the frequencies $\omegtwosqd$, 
$\omegfoursqd$, and $\omegfivesqd$
given in equations (\ref{eq:omegatwosqd}), 
(\ref{eq:omegafoursqd}), and (\ref{eq:omegafivesqd}) 
respectively, we can rewrite equation~(\ref{eq:stabcritgen}) as
\begin{equation}
nk_B \frac{dT}{dz} + 
\frac{d p_{\rm cr}}{d z}
+ \frac{1}{8\pi} \frac{dB^2}{dz} > 0,
\label{eq:stabcritgen2}
\end{equation}
where we have dropped the zero subscripts on the equilibrium
quantities. Equation~(\ref{eq:stabcritgen2}) shows that an ``upwardly
decreasing'' temperature, cosmic-ray pressure, or magnetic pressure is
destabilizing.

We next consider the special case, $k_y=0$, which corresponds to 
pure interchanges.  In this  case,
equation~(\ref{eq:ballim}) leads to the two non-trivial solutions,
\begin{equation}
\sigma=\pm\sqrt{-\frac{b}{b_0}},
\end{equation}
where $b$ is what remains of the coefficient $b_2$ at $k_y=0$.  By
inspection we see that the necessary and sufficient condition for
these modes to be stable is~$b>0$.  
For a vanishing cosmic-ray pressure, the condition~$b>0$
reduces to
\begin{equation}
\label{eq:tserk}
-\frac{d\rho}{dz}>\frac{\rho^2g}{\gamgas\pgas+B^2/4\pi},
\end{equation}
where we have again dropped the zero subscripts on the equilibrium
quantities. Equation~(\ref{eq:tserk}) is the result of
Tserkovnikov~(1960) for the pure interchanges as quoted by
Newcomb~(1961).  The criterion $\dubyasqd>0$ obtained above for the
case $k_y\ne0$ is more restrictive than the condition~$b>0$,
since~$\gamma > 1$.  Thus, $\dubyasqd>0$ is the necessary and
sufficient condition for the plasma to be stable to all modes in the
quasi-interchange limit, including those with~$k_y=0$.

\subsection{Eigenmodes near marginal stability}
\label{sec:ems} 

In this section we consider the properties of unstable
modes very near to the limit of marginal stability. We assume that
$k_y\neq 0$, but take the limit~$k_y H \ll 1$---that is, the parallel
wave length is much longer than the scale height.  Near marginal
stability, the quantity~$b_6$ in equation~(\ref{eq:ballim}) approaches
zero.  There thus exists a solution to the dispersion relation in
which~$\sigma$ also approaches zero, for which the terms proportional
to $\sigma^2$ through~$\sigma^6$ in equation~(\ref{eq:ballim}) can be
neglected.  This solution satisfies the approximate equation
\begin{equation}
\sigma \simeq - \frac{b_6}{b_5}.
\label{eq:dispappr} 
\end{equation} 
Making use of the fact that~$\sigma \rightarrow 0$ for this
mode, we can return to the results of section~\ref{sec:disprel}  
(in particular, the equation ${\mathsf M}\cdot\delta\vvec=0$)
and show that to leading order in~$k_y H$
\begin{equation}
\frac{k_x \delta v_x}{k_y \delta v_y} \simeq - \frac{i k_z p}{\rho g},
\end{equation} 
and
\begin{equation}
\frac{k_z \delta v_z}{k_y \delta v_y} \simeq  \frac{i k_z p}{\rho g},
\end{equation} 
so that
\begin{equation}
|k_x \delta v_x + k_z \delta v_z| \ll |k_y \delta v_y|.
\label{eq:compr} 
\end{equation} 
Thus, for modes with $|k_y H| \ll 1$ near marginal stability, most of
the compression or expansion of the plasma occurs in the direction of
the magnetic field rather than perpendicular to the magnetic field,
despite the fact that~$k_y$ is very small.  We discuss the importance
of this result further in the next section.

\section{Heuristic derivation of stability criterion}
\label{sec:physical} 

\begin{figure}[h]
\includegraphics[width=2.5in]{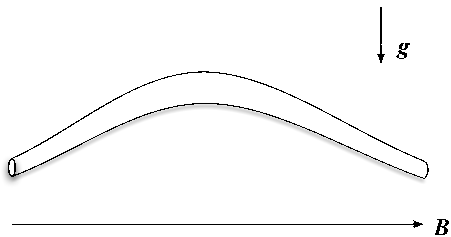}
\caption{\footnotesize An upwardly displaced flux tube.
\label{fig:f1}}
\end{figure}

In this section, we present a way of understanding the stability
criterion in equation~(\ref{eq:stabcrit0}) in physical terms.  We
consider the same equilibrium discussed in section~\ref{sec:disprel} ,
in which~$\gvec = -g \zhat$ and $\Bvec_0 = B_0 \yhat$, and we again
take the plasma to be perfectly conducting, so that magnetic field
lines are frozen-in to the fluid.  However, we now assume that the
equilibrium is very close to marginal stability.  We then imagine some
mode in the plasma that causes a long and narrow magnetic flux tube to
rise upwards, as depicted in Figure~\ref{fig:f1}. For simplicity, we
assume that the ends of the flux tube are anchored at the flux tube's
initial height. We take the flux tube to be very long, so that
magnetic tension forces are very weak.  Because the medium is
arbitrarily close to marginal stability, the growth time or
oscillation time for the mode is arbitrarily long. Thus, even though
the flux tube is long, there is plenty of time for conduction and
diffusion to equalize~$T$ and $p_{\rm cr}$ along the perturbed
magnetic field lines. We assume that the total pressure, $p_{\rm tot}
= \pgas + \pcosmic + B^2/8\pi$, at each point along the flux tube is
equal to the total pressure just outside the flux tube at that
point.\footnote{Total-pressure variations are associated with
  high-frequency magnetosonic waves. These waves are stable at $\beta
  \gg 1$ \paperi, and we assume they are stable here
  as well. However, we note that Socrates, Parrish, \& Stone (2007)
  have shown that magnetosonic waves can become unstable in the
  presence of an equilibrium heat flux, when $B_{0z} \neq 0$.}

We define $\Delta n$, $\Delta T$, $\Delta B^2$ and $\Delta p_{\rm
  cr}$, respectively, as the difference between the density,
temperature, field-strength-squared, and cosmic-ray pressure at the
highest point in our flux tube and the immediately surrounding medium,
at a point in time when the top of the flux tube is a small
distance~$\Delta z$ above the flux tube's initial height. The
constancy of~$T$ and $p_{\rm cr}$ along the flux tube yields the
relations (accurate to first-order in~$\Delta z/H$)
\begin{equation}
\Delta T = - \Delta z \frac{dT}{dz}
\label{eq:dt1} 
\end{equation} 
and
\begin{equation}
\Delta p_{\rm cr} = - \Delta z \frac{dp_{\rm cr}}{dz},
\label{eq:dpcr1} 
\end{equation} 
where $dT/dz$ and $dp_{\rm cr}/dz$ are the gradients of the
equilibrium temperature and cosmic-ray pressure evaluated at the
initial height of the flux tube. 

Equations~(\ref{eq:dt1}) and (\ref{eq:dpcr1}) tell us how to evaluate
$T$ and $p_{\rm cr}$ in our flux tube. Evaluating $B^2$ in the flux
tube is a little more involved.  Assuming that the total pressure
decreases with height, the fluid in the flux tube has to expand in
order to achieve total-pressure balance.  However, the manner in which
the flux tube expands is not obvious. If the plasma expands primarily
along the magnetic field, the cross sectional area of the flux tube
will be constant along the flux tube, and thus so will the magnetic
field strength. On the other hand, if the plasma expands perpendicular
to the field, the magnetic field strength will decrease. Which type of
expansion does the plasma favor?  We answer this question analytically
in section~\ref{sec:ems}, where we show that, near marginal stability,
$|k_y \delta v_y| \gg |k_x \delta v_x + k_z \delta v_z|$ for the
low-frequency long-parallel-wavelength buoyancy instability in the
large-$|k_x|$ limit. Thus, for this mode, most of the expansion
$\nabla \cdot \vvec$ arises from the parallel motion. We note that
this statement is stronger than the statement that $|v_y| \gg |v_x|$,
because we take $|k_x| \gg |k_y|$.

How can we understand this result in physical terms?  One way is by
analogy to the $\delta W$ analysis of the stability of ideal MHD
plasmas, in the absence of thermal conduction. (Bernstein, Frieman,
Kruskal, \& Kulsrud 1958, Friedberg~1987) In this analysis, it is
shown that if a mode expands in the direction perpendicular to the
magnetic field, additional work must be done on the surrounding
magnetic field. This requirement makes the mode more stable. To find
the stability criterion, we must seek out the most unstable mode,
which in this case is a mode that keeps the cross-sectional area of
the flux tube constant.

Taking the cross-sectional area of the flux tube to be constant,
we can treat $B^2$ as constant along the flux tube. This allows
us to write that
\begin{equation}
\Delta B^2 = - \Delta z \frac{dB^2}{dz}.
\label{eq:dB21} 
\end{equation} 
The condition that the total pressure inside the flux tube matches the
total pressure outside the flux tube can be written as
\begin{equation}
k_B T \Delta n  + n k_B \Delta T 
+ \Delta p_{\rm cr} + \frac{\Delta B^2}{8\pi} = 0.
\label{eq:peq} 
\end{equation} 
Together, equations~(\ref{eq:dt1}) through (\ref{eq:peq}) imply that
\begin{equation}
k_B T \Delta n = \Delta z\left( n k_B \frac{dT}{dz} + \frac{dp_{\rm cr}}{dz}
+ \frac{1}{8\pi} \frac{dB^2}{dz} \right).
\label{eq:stabcritf} 
\end{equation} 
The stability criterion, equation~(\ref{eq:stabcrit0}), is thus the
condition that the material inside an upwardly displaced,
long, and narrow flux tube be denser than the surrounding
medium.

\section{Convection in galaxy cluster plasmas}
\label{sec:gc}

In many galaxy-cluster cores, the radiative cooling time is much
shorter than the cluster's likely age (Fabian 1994). Nevertheless,
high-spectral-resolution X-ray observations show that very little
plasma actually cools to low temperatures. (B\"{o}hringer et~al~2001;
David et al 2001; Molendi \& Pizzolato~2001; Peterson et~al~2001,
2003; Tamura et~al~2001; Blanton, Sarazin, \& McNamara 2003). This
finding, sometimes referred to as the ``cooling-flow problem,''
strongly suggests that plasma heating approximately balances radiative
cooling in cluster cores.

A heating mechanism for cluster cores
that has been studied extensively is heating by a
central active galactic nucleus (AGN). The importance of such ``AGN
feedback'' is suggested by the observation that almost all clusters
with strongly cooling cores possess active central radio sources
(Burns 1990; Ball, Burns, \& Loken 1993; Eilek~2004) and by the
correlation between the X-ray luminosity from within a cluster's
cooling radius and the mechanical luminosity of a cluster's central
AGN (B\^irzan et al 2004,~Eilek 2004).  One of the main unsolved
problems regarding AGN feedback is to understand how AGN power is
transferred to the diffuse ambient plasma.  A number of mechanisms
have been investigated, including Compton heating (Binney \& Tabor
1995; Ciotti \& Ostriker 1997, 2001; Ciotti, Ostriker, \&
Pellegrini~2004, Sazonov et al 2005), shocks (Tabor \& Binney 1993,
Binney \& Tabor 1995), magnetohydrodynamic (MHD) wave-mediated plasma
heating by cosmic rays (B\"{o}hringer \& Morfill~1988; Rosner \&
Tucker~1989; Loewenstein, Zweibel, \& Begelman~1991), and cosmic-ray
bubbles produced by the central AGN (Churazov et al~2001, 2002;
Reynolds 2002; Br\"{u}ggen~2003; Reynolds et~al~2005), which can heat
intracluster plasma by generating turbulence (Loewenstein \& Fabian
1990, Churazov et~al~2004, Cattaneo \& Teyssier 2007) and sound waves
(Fabian et~al~2003; Ruszkowski, Br\"{u}ggen, \& Begelman 2004a,b) and
by doing $pdV$ work (Begelman 2001, 2002; Ruszkowski \& Begelman~2002;
Hoeft \& Br\"{u}ggen~2004).

Another way in which central AGNs may heat the
intracluster medium is by accelerating cosmic rays that mix
with the intracluster plasma and cause the
intracluster medium to become convectively unstable.  A steady-state,
spherically symmetric, mixing-length model based on this idea was
developed by Chandran (2004) and subsequently refined by Chandran
(2005) and Chandran \& Rasera (2007).  In this model, a central
supermassive black hole accretes hot intracluster plasma at the Bondi
rate, and converts a small fraction of the accreted rest-mass energy
into cosmic rays that are accelerated by shocks within some
distance~$r_{\rm source}$ of the center of the cluster.  The resulting
cosmic-ray pressure gradient leads to convection, which in turn heats
the thermal plasma in the cluster core by advecting internal energy
inwards and allowing the cosmic rays to do~$pdV$ work on the thermal
plasma. The model also includes thermal conduction, cosmic-ray
diffusion, and radiative cooling.  By adjusting a single parameter in
the model ($r_{\rm source}$), Chandran
\& Rasera (2007) were able to achieve a good match to the observed
density and temperature profiles in a sample of eight clusters.

The treatment of convective stability in the work of
Chandran (2004, 2005) and Chandran \& Rasera~(2007) was based on the
assumption that~$\beta = 8\pi p/B^2 \gg 1$. The 
present paper investigates convective stability for arbitrary~$\beta$. One
of the motivations for this work is the possibility
that some clusters with short central cooling times (``cooling-core
clusters'') may
be in the $\beta \sim 1$ regime. For a fully ionized plasma with a
hydrogen mass fraction~$\,X=0.7$ and helium mass fraction $\,Y=0.29$,
\begin{equation}
\beta = 6.3 \times \left(\frac{n_e}{10^{-2} \mbox{cm}^{-3}}\right) \times
\left(\frac{k_B T}{\mbox{3 keV}}\right) \times
\left(\frac{B}{10 \mu \mbox{G}}\right)^{-2}.
\label{eq:beta} 
\end{equation} 
Although many studies of the magnetic field strength in clusters of
galaxies find $B$ in the range of $1-5 \mu$G (see, e.g., Kronberg
1994, Eilek \& Owen 2002), some studies of Faraday rotation in
cooling-core clusters find much stronger magnetic fields (Taylor \&
Perley 1993; Kronberg 1994; Taylor, Fabian, \& Allen 2002).  In the
case of Hydra~A, Taylor \& Perley~(1993) found a tangled magnetic
field of~$\sim 30\mu$G, and Taylor, Fabian, \& Allen~(2002) found a
tangled magnetic field of $\sim 35\mu$G.  The analysis of X-ray
observations of Hydra~A carried out by Kaastra et al (2004), when
converted to a $\Lambda$CDM cosmology (see Chandran \& Rasera 2007),
indicate that $n_e \simeq 0.01 \mbox{ cm}^{-3}$ and $k_B T \simeq
3.4$~keV in Hydra~A at $r=50$~kpc.  Equation~(\ref{eq:beta}) thus
shows that if $B$ is indeed as large as~$30 \mu$G in the core of
Hydra~A, then $\beta$ is of order unity. Values of $\beta \sim 0.1-1$
for cluster cores in several other galaxy clusters were reported by
Eilek \& Owen~(2002).  Although these studies suggest that $\beta \sim
1$ magnetic fields could be common in cooling-core clusters, some
caution is warranted here. Vogt \& Ensslin~(2005) have reanalyzed the
Faraday-rotation data for Hydra~A using an updated plasma
density profile, and found an rms magnetic field of~$7 \mu$G, which
corresponds to~$\beta =15$ at~$r=50$~kpc in Hydra~A.
In the remainder of this section, we explore the implications of
the condition~$\beta \lesssim 1$ on convective instability in clusters,
but the above uncertainty in the value of~$\beta$ in cooling-core
clusters should be born in mind.

In section~\ref{sec:qil} we showed that the
necessary and sufficient condition for stability for a mode with fixed
nonzero~$k_y$ in the quasi-interchange limit ($|k_x|$ much larger than
$|k_y|$, $|k_z|$, and $H^{-1}$) is
\begin{equation}
k_y^2 v_A^2 + g \left(\frac{d\ln T}{dz}
+ \frac{\pcosmic}{\pgas}\frac{d\ln \pcosmic}{dz}
+ \frac{B^2}{8\pi \pgas}\frac{d\ln B^2}{dz}
\right) > 0.
\label{eq:stabcrit2} 
\end{equation} 
This equation shows that the magnetic field has two competing effects
on convective stability.  First, if the field strength decreases
``upwards'' (i.e., $dB^2/dz < 0$), the $g\beta^{-1}\, d\ln B^2/dz$
``magnetic-buoyancy term'' in equation~(\ref{eq:stabcrit2}) is
destabilizing. On the other hand, the $k_y^2 v_A^2$ ``magnetic-tension
term'' is stabilizing.  We can estimate the relative importance of the
different terms in equation~(\ref{eq:stabcrit2}) by defining the
length scales~$H_f$, $H_B$, and $H_p$ via the equations
\begin{equation}
H_f^{-1} = \left|\frac{d\ln T}{dz} +
\frac{\pcosmic}{\pgas}\frac{d\ln\pcosmic}{dz}\right|,
\end{equation}
\begin{equation}
H_B^{-1} = \left|\frac{d\ln B}{dz}\right|,
\end{equation} 
and
\begin{equation}
H_p^{-1} = \frac{\rho g}{p}.
\end{equation} 
The ratio of the magnetic-tension term to the magnetic-buoyancy 
term is then
\begin{equation}
\frac{k_y^2 v_A^2}{2g\beta^{-1} H_B^{-1}} = k_y^2 H_B H_p,
\label{eq:ratio1} 
\end{equation} 
while the ratio of the magnetic-tension term to the 
``fluid terms,'' $g[d\ln T/dz + (\pcosmic/\pgas)d\ln \pcosmic/dz]$ is
\begin{equation}
\frac{k_y^2v_A^2}{g H_f^{-1}} = 2\beta^{-1} k_y^2 H_p H_f.
\label{eq:ratio2} 
\end{equation} 
The magnetic field turns off the buoyancy instability at wavevectors
for which the magnetic tension term dominates over both the magnetic
buoyancy term and the fluid terms.  At $\beta\sim 1$, this happens for
$k_y^2 H_p H_B \gg 1$ and $k_y^2 H_p H_f\gg 1$.  If we take all the
scale lengths to be comparable to the density scale height~$H$, then
at $\beta \sim 1$ magnetic tension turns off the instability for~$k_yH
\gg 1$, but is negligible for $k_yH \ll 1$.  At $\beta\sim 1$ and $k_y
H \sim 1$, the tension, buoyancy, and fluid terms are all comparable,
and magnetic buoyancy and magnetic tension to some extent cancel out.
When $\beta \ll 1$, magnetic tension dominates for~$k_y H \gg 1$,
magnetic buoyancy dominates for $k_y H \ll 1$, and the two are
comparable at $k_y H \sim 1$, again assuming that $H_p \sim H_B \sim
H_f \sim H$.

To apply our results to galaxy-cluster plasmas, we imagine some
hypothetical spherical equilibrium, and consider local modes at a
radius~$r$ at some location where the radial component of the magnetic
field vanishes, and where all the scale lengths are of order~$r$.  Our
local analysis of a slab-symmetric equilibrium is strictly applicable
only to modes with $k_y r \gg 1$, that is, to modes with parallel
wavelengths much less than the scale height.  Our results show that
such modes are stable when~$\beta \lesssim 1$ because of the stabilizing
effects of magnetic tension. 

\section{The Parker instability in the interstellar medium}
\label{sec:parker}

The Parker instability is an unstable mode in a gravitationally
stratified plasma that is driven by the buoyancy of the magnetic field
and/or cosmic rays. (Parker 1966, 1967) The Parker instability is
thought to be important for the interstellar medium for several
reasons.  It has been argued that this mode, acting alone or in
concert with the thermal instability (Field 1965), contributes to the
formation of molecular clouds (Blitz \& Shu 1980; Parker \& Jokipii
2000; Kosinski \& Hanasz 2005, 2006, 2007). It has also been suggested that 
the Parker instability is a mechanism for regulating the transport of
magnetic fields and cosmic rays in the direction perpendicular to the
galactic plane, and for driving the Galactic dynamo.  (See, e.g.,
Parker 1992, Hanasz \& Lesch 2000, Hanasz et~al~2004).

The Parker instability is very similar to the instability that we have
investigated in this paper. Standard analyses of the Parker
instability consider an equilibrium in which $\gvec = - g\zhat$,
$\Bvec$ is in the $xy$-plane, $\rho \propto \exp(-z/H)$, and $H$, $T$,
$\beta$, and $p_{\rm cr}/p$ are constant. (Parker 1966, 1967; Shu
1974, Ryu et~al~2003). In early studies, the parallel cosmic-ray
diffusion coefficient~$D_\parallel$ was taken to be infinite, since
$p_{\rm cr}$ was assumed to be constant along magnetic field
lines. (Parker 1966, 1967; Shu 1974) On the other hand, Ryu
et~al~(2003) considered the effects of finite~$D_\parallel$, as well
as cosmic-ray diffusion perpendicular to magnetic field lines.  All of
these studies took the thermal plasma to be adiabatic.

The analysis of the present paper extends our understanding of the
Parker instability in two ways. First, we allow the equilibrium
values of $T$, $\beta$, 
and $p_{\rm cr}/p$ to vary with~$z$. Second, we consider the effects of
anisotropic thermal conduction. By doing so, we show that the
condition~$dT/dz<0$ makes a plasma more unstable to the Parker
instability than when the equilibrium is isothermal.
We also show that even if the equilibrium is isothermal,
anisotropic thermal conduction makes a
stratified plasma more unstable to the Parker instability than when
the plasma is treated as adiabatic. The Parker stability criterion
in the limit~$|k_x| \rightarrow \infty$
for the equilibrium described above can be obtained from
the $k_y \rightarrow 0$ limit of equation~(73)
of Shu~(1974):
\begin{equation}
\frac{B^2}{8\pi} + p_{\rm cr} < (\gamma-1) p .
\label{eq:parker1} 
\end{equation} 
Multiplying this equation by~$-1/H$ and making use of the assumptions
that $dB^2/dz = - B^2/H$,
$dp_{\rm cr}/dz = - p_{\rm cr}/H$, and $dp/dz = - p/H$,
we can rewrite equation~(\ref{eq:parker1}) as
\begin{equation}
\frac{d}{dz} \left(\frac{B^2}{8\pi} + p_{\rm cr}\right) > -\frac{(\gamma-1)p}{H}.
\label{eq:parker2} 
\end{equation} 
On the other hand, when anisotropic thermal conduction is taken into
account, the stability criterion for this constant-temperature
equilibrium from equation~(\ref{eq:stabcritgen2}) is
\begin{equation}
\frac{d}{dz} \left(\frac{B^2}{8\pi} + p_{\rm cr}\right) > 0.
\label{eq:parker3} 
\end{equation} 
Since $\gamma >1$, equation~(\ref{eq:parker3}) is more restrictive
than equation~(\ref{eq:parker2}), and anisotropic thermal conduction
allows for instability under a larger range of equilibria than when
the plasma is taken to be adiabatic.  The reason for this is that as a
fluid parcel rises and expands, anisotropic thermal conduction allows
heat to flow up along the magnetic field lines into the rising fluid
parcel. This heat flow increases the temperature of the rising fluid
parcel relative to the adiabatic case and thereby lowers the density,
making the fluid parcel more buoyant, as in the high-$\beta$
zero-$p_{\rm cr}$ limit considered by Balbus~(2000, 2001).

\section{Conclusion}
\label{sec:conc}

In this paper we derive the stability criterion for local buoyancy
instabilities in a stratified plasma, with the equilibrium magnetic
field in the~$\yhat$ direction and gravity in the~$-\zhat$ direction.
We take into account cosmic-ray diffusion and thermal conduction along
magnetic field lines and focus on the large-$|k_x|$ limit, which is
the most unstable limit for adiabatic plasmas.  Our work extends the
earlier work of Balbus (2000, 2001) and \paperi\ by allowing for
arbitrarily strong magnetic fields.  Applying our work to
galaxy-cluster plasmas, we find that increasing the magnetic field to
the point that~$\beta = 8\pi p/B^2 \lesssim 1$ would shut off buoyancy
instabilities at wavelengths along the magnetic field that are much
shorter than the equilibrium scale height.  Our analysis also extends
our understanding of the Parker instability by allowing the
equilibrium values of $T$, $\beta$, and $p_{\rm cr}/p$ to vary
with~$z$, and by accounting for anisotropic thermal conduction. We
find that the interstellar medium is more unstable to the Parker
instability than was predicted by earlier studies, which treated the
thermal plasma as adiabatic.

\acknowledgements
We thank Eliot Quataert for helpful discussions.
This work was partially supported by NASA's Astrophysical 
Theory Program under grant NNG 05GH39G and by NSF under
grant AST 05-49577

\appendix
\section{Approximate analytical solutions 
            and comparisons to numerical 
            solutions}
\label{ap:eigenmodes} 

In this appendix, in order to provide further insight into 
buoyancy instabilities in the equilibrium described in
section~\ref{sec:disprel}, we derive a set of approximate analytical
solutions to equation~(\ref{eq:ballim}). To check our analytical
solutions, we compare them with numerical solutions of the general
dispersion relation [equation~(\ref{eq:gendisp})]. We begin by
defining the quantities~$\omegrefsqd$~and~$\kcrit$:
\begin{equation}
\omegrefsqd=\frac{p_0}{\rho_0H^2},
\end{equation}
and
\begin{equation}
\kcrit=\sqrt{\frac{\omegref}{D_{\rm cond}}}.
\end{equation}
We assume that $D_{\rm cond}\sim D_\parallel$, and that
the equilibrium cosmic-ray pressure is not very small
compared to the equilibrium thermal pressure. The 
frequency~$\omegref$ is comparable to the 
frequency of buoyancy oscillations in the medium except when 
the plasma is near the stability boundary for one 
of the buoyancy instabilities.  The quantity~$\kcrit$ is defined 
so that when $k_y = \kcrit$ we have $\omegref = \eta \sim \nu$.
These definitions and observations allow us to define two 
limits for which it is possible to derive approximate 
analytical solutions to equation~(\ref{eq:ballim}); the 
``long-parallel-wavelength'' limit 
($\left|k_y\right|\ll \kcrit$) for which the diffusive 
frequencies $\nu$ and $\eta$ are small compared to the 
buoyancy frequencies, and the ``short-parallel-wavelength'' 
limit ($\left|k_y\right|\gg \kcrit$) for which the diffusive 
frequencies are large compared to the buoyancy frequencies. 
In the following subsections, we consider each of these limits 
in turn, and as a further check we also compare their 
high$-$$\beta$ limits with the results of 
\paperi.

\subsubsection{Long-parallel-wavelength limit}
               
As stated above, in the long-parallel-wavelength limit we 
have $\eta, \nu \ll \omegrefsqd$, while all of the other 
quantities listed in
 equations~(\ref{eq:omegassqd})$-$(\ref{eq:omegafivesqd}) 
that remain in the dispersion relation 
[equation~(\ref{eq:ballim})] are $\sim\omegrefsqd$. We 
take advantage of this and apply the method of dominant 
balance (Bender \&  Orszag, 1978). We set
$\sigma=\omegref\tilde\sigma$, 
$\omegzero=\omegref\tomegzero$,
$\omegone=\omegref\tomegone$,
$\omegtwo=\omegref\tomegtwo$,
$\omegthree=\omegref\tomegthree$,
$\omegfour=\omegref\tomegfour$,
$\omegfive=\omegref\tomegfive$,
$\omegalf=\omegref\tomegalf$,
$\eta=\epsilon\omegref\teta$,
$\nu=\epsilon\omegref\tnu$,
where $\epsilon\ll1$,
and substitute these into equation~(\ref{eq:ballim}) 
leading to the result:
\begin{equation}
\label{eq:nodimb}
\tilde b_0\tilde\sigma^6+
\tilde b_1\tilde\sigma^5+
\tilde b_2\tilde\sigma^4+
\tilde b_3\tilde\sigma^3 +
\tilde b_4\tilde\sigma^2+
\tilde b_5\tilde\sigma +
\tilde b_6 =0,
\end{equation}
where the coefficients are now given by:
\begin{eqnarray}
\tilde b_0&=&\gamgas+\chi\gamcos+\twooverbeta,\\
\tilde b_1&=&
\epsilon\left[
 \tilde\nu\left(\gamgas+\twooverbeta\right)+\tilde \eta\left(1+\chi\gamcos+\twooverbeta\right)
 \right],\\
\tilde b_2&=&\left(\gamgas+\chi\gamcos\right)\left(\tomegalfsqd+
\tomegzerosqd+\tcrit\right)-\tomegzerosqd+\twooverbeta\left(\tomegzerosqd+\tcrit\right)+
\epsilon^2\tilde\nu\tilde\eta\left(1+\twooverbeta\right),\\
\tilde b_3&=&\epsilon
\Biggl\{
\left[
\tilde\eta\left(1+\chi\gamcos\right)+\tilde\nu\gamgas
\right]
\left(\tomegalfsqd+\tomegzerosqd+\tcrit\right)+
\left(\tilde\nu+\tilde\eta\right)
\left[
\twooverbeta\left(\tomegzerosqd+\tcrit\right)-\tomegzerosqd
\right]
\Biggr\},\qquad\quad
\end{eqnarray}
\begin{eqnarray}
\tilde b_4&=&\tomegalfsqd
\left[
\left(\gamgas+\chi\gamcos\right)\left(\tomegzerosqd+\tcrit\right)-\tomegzerosqd
\right]
+\epsilon^2\tilde\nu\tilde\eta
\left[
\tomegalfsqd+\crit+\twooverbeta\left(\tomegzerosqd+\tcrit\right)
\right], \qquad\qquad\\
\tilde b_5&=&
\epsilon\,\tomegalfsqd\left[
\left(\tomegzerosqd+\tcrit\right)
\left[
\tilde\nu\gamgas+\tilde\eta\left(1+\chi\gamcos\right)
\right]
-\left(\tilde\nu+\tilde\eta\right)\tomegzerosqd
\right]
, \\
\tilde b_6&=&\epsilon^2\tilde\nu\tilde\eta\tomegalfsqd\tcrit\label{eq:nodimbcoff6}. 
\end{eqnarray}
We may now solve for the two sets of approximate solutions 
to equation~(\ref{eq:nodimb}) by assuming that one set will 
satisfy $\sigma\sim\omegref$ while the other satisfies 
$\sigma\sim\epsilon\omegref$.  We present each of these 
solutions in turn in the following subsections.

\subsubsubsection{Adiabatic-buoyancy modes}
We refer to the set of solutions satisfying $\sigma\sim\omegref$ as the 
``adiabatic-buoyancy modes'' since these are
of the order of the buoyancy frequencies except when near 
the limit of marginal stability.
\footnote{We note for clarity that in the limiting case of 
high-$\beta$ discussed in \paperi, these modes were referred to 
as the ``adiabatic convective/buoyancy modes.''}
To obtain expressions for these modes we set,
\begin{equation}
\tsigma=\tsigma_0+\epsilon\tsigma_1+\ldots,
\end{equation}
where $\tsigma_0$ represents the lowest-order part of $\tsigma$
and $\tsigma_1$ is the first-order correction.
Substituting this expansion into equation~(\ref{eq:nodimb}),
collecting terms that are of like-order in $\epsilon$,
and requiring sums of terms of like-order in $\epsilon$ to vanish 
separately we find for the lowest-order terms:
\begin{equation}
\label{eq:ballong}
\tilde a\tilde\sigma_0^4+\tilde b\tilde\sigma_0^2+\tilde c=0,
\end{equation}
where
\begin{eqnarray}
\tilde a &=& \gamgas+\chi\gamcos+\twooverbeta, \\
\tilde b &=& \left(\gamgas+\chi\gamcos+\twooverbeta\right)\tcrit
+
\left[
\left(\gamgas-1\right)+\chi\gamcos+\twooverbeta
\right]\tomegzerosqd +
\left(\gamgas+\chi\gamcos\right)\tomegalfsqd,\qquad
\\
\tilde c &=&\tomegalfsqd
\left\{
\left(\gamgas+\chi\gamcos\right)\tcrit+
\left[
\left(\gamgas-1\right)+\chi\gamcos
\right]\tomegzerosqd
\right\}.
\end{eqnarray}
Restoring the dimensions we write the solutions with the 
notation:
\begin{equation}
\label{eq:dimrest}
\sigma_{0,\pm\pm} \simeq \pm\sqrt{\frac{-b\pm\sqrt{b^2-4ac}}{2a}},
\end{equation}
where $a$, $b$, and $c$ are the dimensional analogs of 
$\tilde a$, $\tilde b$, and $\tilde c$ respectively,
and where the {left-most} $\pm$ subscript on $\sigma$ shall 
refer to the $\pm$ symbol {inside} the radical while the 
{right-most} refers to the $\pm$ symbol {outside} the 
radical. We note that the solution $\sigma_{0,++}$ is 
unstable when $c<0$. More explicitly,
\begin{equation}
\label{eq:unstableAB}
\left(\gamgas+\chi\gamcos\right)\crit+
\left[
\left(\gamgas-1\right)+\chi\gamcos
\right]\omegzerosqd<0,
\end{equation}
which holds only when $\crit$ is sufficiently negative. We 
define the buoyancy frequency $N$ through the equation
\begin{equation}
\label{eq:defN}
N^2=\left[
\gamgas\omegonesqd+\chi\gamcos\omegthreesqd+\oneoverbeta\omegfivesqd
\right]
\left[
\gamgas+\chi\gamcos+\twooverbeta
\right]^{-1},
\end{equation}
which is related to $\dubyasqd$ through the identity,
\begin{equation}
\label{eq:dubyaNident}
\dubyasqd+\frac{\left[\left(\gamgas-1\right)+\chi\gamcos \right] }
{\gamgas+\chi\gamcos}\omegzerosqd = \frac{\left(\gamgas+\chi\gamcos+2/\beta\right)}
{\gamgas+\chi\gamcos}N^2.
\end{equation}
In a high$-$$\beta$ plasma in the absence of cosmic rays,
$N$ reduces to the Brunt-V\"ais\"al\"a frequency for 
buoyancy oscillations in a gravitationally-stratified 
medium. With $N$ defined as in equation~(\ref{eq:defN}) 
we may rewrite the condition in 
equation~(\ref{eq:unstableAB}) as
\begin{equation}
\frac{\omegalfsqd}{1+2/\left[\beta\left(\gamgas+\chi\gamcos\right)\right]}+N^2<0,
\end{equation}
which, in the limit of high$-$$\beta$, reduces to the 
corresponding result obtained previously for these modes in 
\paperi.
 \begin{figure*}[ht]
 \hfill
 \includegraphics[width=6.0in, keepaspectratio=true,trim= 0 0 0 0 ]{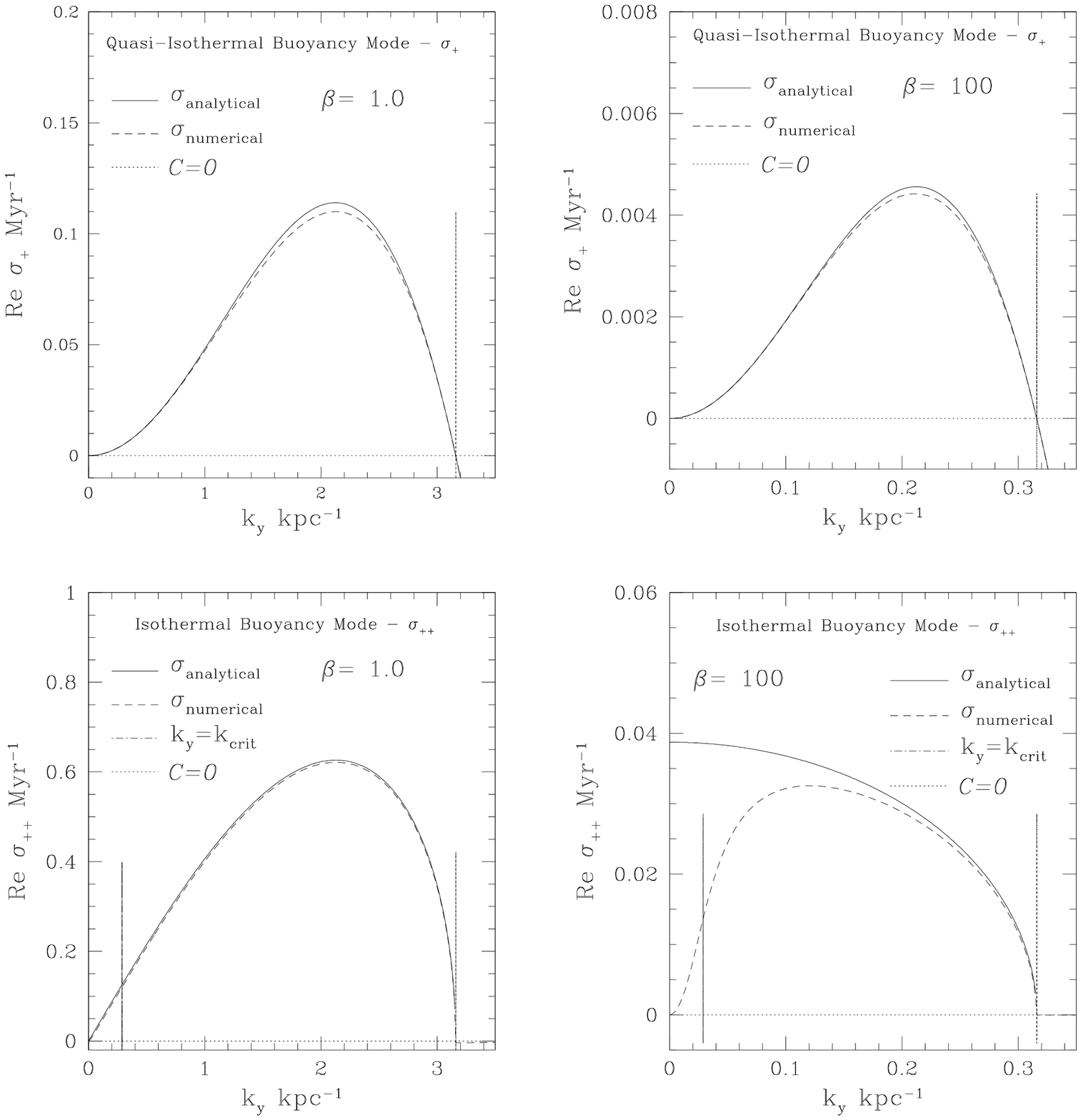}
 \hfill\ 
\caption {\footnotesize Unstable solutions of the dispersion relation for the long-parallel-wavelength 
          limit (first row) and the short-parallel-wavelength limit (second row) for the cases $\beta=1$
          (left panels) and $\beta=100$ (right panels).  The solid curves correspond to the approximate 
          analytical solutions given in the text, the dashed curves correspond to numerical solutions of 
          equation~(\ref{eq:gendisp}), the intesection of the dotted lines indicate the location where 
          $\crit=0$, and in the bottom row, the dashed-dotted vertical line shows the location where 
          $k_y=k_{\rm crit}$. (for the long-parallel-wavelength limit in the top row, $\kcrit$ falls 
          outside the range of wave numbers shown.) 
 \label{fig:f2}}
 \end{figure*}

\subsubsubsection{Quasi-isothermal buoyancy modes}
We refer to the set of solutions which satisfy $\sigma\sim\epsilon\omegref$
as the ``quasi-isothermal buoyancy modes,'' since these modes
are of the same order as the frequencies $\eta$ and $\nu$ in the
long-parallel-wavelength limit. To obtain the quasi-isothermal 
buoyancy modes, we now set $\tsigma_0 = 0$, so that
$\tsigma = \epsilon\tsigma_1+O(\epsilon^2)$. Substituting 
into equations~(\ref{eq:nodimb})$-$(\ref{eq:nodimbcoff6}) 
and retaining only lowest-order terms we find:
\begin{equation}
\label{eq:qisothermal}
a\sigma_1^2+b\sigma_1+c=0,
\end{equation}
where now,
\begin{eqnarray}
a&=& \left(\gamgas+\chi\gamcos\right)\crit+
\left[
\left(\gamgas-1\right)+\chi\gamcos
\right]\omegzerosqd, 
\label{eq:itcoffa} \\
b&=& f\crit+K\omegzerosqd,
\label{eq:itcoffb} \\
c&=&\nu\eta\crit,
\label{eq:itcoffc}
\end{eqnarray}
and where the definition of $K$ is given in 
equation~(\ref{eq:defK}) and we have introduced 
the positive definite quantity:
\begin{equation}
f = \gamgas\nu+\eta\left(1+\chi\gamcos\right).
\end{equation}
For notational convenience we have again reverted to the 
dimensional form of our solution.  The discriminant of 
equation~(\ref{eq:qisothermal}) can be expressed in the 
form,
\begin{equation}
b^2-4ac=\left\{
\left[
\gamgas\nu-\eta\left(1+\chi\gamcos\right)
\right]\crit+
\left[
\left(\gamgas-1\right)\nu-\chi\gamcos\eta
\right]\omegzerosqd
\right\}^2 +
4\eta\nu\left(\gamgas-1\right)\chi\gamcos\left(\crit+\omegzerosqd\right)^2,
\end{equation}
which is non-negative so that the solutions,
\begin{equation}
\sigma_1 = \frac{-b\pm \sqrt{{b^2-4ac}}}{2a},
\end{equation}
are both real.

To examine the stability of these solutions we first note
that when all three coefficients of 
equation~(\ref{eq:qisothermal}) are non-zero, both roots 
will be negative if and only if the coefficients also all 
have the same sign. The necessary and sufficient condition 
for this is $ac>0$. Inspection of 
equations~(\ref{eq:itcoffa})$-$(\ref{eq:itcoffc}) indicate 
that if $c$ is positive then both $a$ and $b$ are also 
positive.
If instead, $c<0$, then we must have $a<0$ for both roots to 
have the same sign, in which case $b<0$ as well and the 
resulting roots are again both negative. Thus an unstable 
mode results only when $c<0<a$, or explicitly:
\begin{equation}
\label{eq:unstableQI}
\nu\eta\crit<0<\left(\gamgas+\chi\gamcos\right)\crit+
\left[
\left(\gamgas-1\right)+\chi\gamcos
\right]\omegzerosqd.
\end{equation}
Once again making use of equation~(\ref{eq:dubyaNident}) we 
may rewrite this result as 
\begin{equation}
\omegalfsqd+\dubyasqd<0<\frac
{\omegalfsqd}{1+2/\left[\beta\left(\gamgas+\chi\gamcos\right)\right]}+N^2,
\end{equation}
an expression which again reduces in the high$-$$\beta$ 
limit to the result obtained for these modes in \paperi\
where they are referred to as the ``quasi-isothermal convective'' 
modes.  

Note that when both inequalities in 
equation~(\ref{eq:unstableQI}) are satisfied, the quasi-isothermal 
mode is unstable, while if only the first inequality is satisfied,
the quasi-isothermal mode is stable and the adiabatic buoyancy mode
is unstable so that we are guaranteed an unstable mode whenever
$\crit<0$. Since these results must hold for arbitrarily small
finite values of $k_y$, we conclude that these inequalities are 
consistent with our stability criterion $\dubyasqd>0$.

\subsubsubsection{Comparison with numerical solutions}
\label{sec:numcomplong}
As a final check of the solutions, we present numerical
solutions to the general dispersion relation
[equation~(\ref{eq:gendisp})], using a suitably chosen set of
parameters for comparison with the predictions of
equation~(\ref{eq:dimrest}). For the particular set of
parameters chosen, the unstable mode is the quasi-isothermal
buoyancy mode. Results for this mode are presented for the
unstable quasi-isothermal buoyancy mode in the top row of figure
\ref{fig:f2} for the cases $\beta=1$ and $\beta=100$.  For the
entire range of wave number shown, we have $k_y\gg \kcrit$,
so we do not indicate the location of $\kcrit$ in the figure.
The analytical solutions in these figures are seen to compare
well with the numerical results, and the growth rates become negative
where the quantity $\crit$ passes from negative to positive values as
indicated by the intersection of the vertical and horizontal dotted
lines, demonstrating that the solutions honor the stability criterion
$\crit > 0$. For the $\beta=1$ case, the mode is unstable in the range
$0 < k_y\lesssim 3.16 \perkpc$, and the maximum growth rate occurs 
where $k_y=k_{y,{\rm max}}\simeq2.1\perkpc$ and takes the value
$\sigma_{+,{\rm max}}\simeq0.114\permyr$, while for the $\beta=100$ case,
the range of unstable wave numbers is $0<k_y\lesssim 0.316\perkpc$,
$k_{y,{\rm max}}\simeq0.21\perkpc$ and $\sigma_{+,{\rm max}}\simeq5.44\times10^{-3}\permyr$.
These results illustrate that a dynamically significant magnetic
field results in a significantly larger maximum rate of growth
for the instability as well as a much larger range of unstable
wave numbers for these modes.

\subsubsection{Short-parallel-wavelength limit}
We now take up the case $|k_y|\gg k_{\rm crit}$ which we
recall is the limit in which the rates of diffusion and
conduction are large relative to the buoyancy frequencies.
Once again  we set 
$\sigma=\omegref\tilde\sigma$,
$\omegzero=\omegref\tomegzero$,
$\omegone=\omegref\tomegone$,
$\omegtwo=\omegref\tomegtwo$,
$\omegthree=\omegref\tomegthree$,
$\omegfour=\omegref\tomegfour$,
$\omegfive=\omegref\tomegfive$,
$\omegalf=\omegref\tomegalf$,
as before; but we scale the diffusive frequencies according
to
$\nu=\epsilon^{-1}\tilde\nu$, and
$\eta=\epsilon^{-1}\tilde\eta$.
and we define the scaled coefficients
$\tilde b_0=b_0$,
$\tilde b_1=b_1\omegref$,
$\tilde b_2=b_2\omegref^2$,
$\tilde b_3=b_3\omegref^3$,
$\tilde b_4=b_4\omegref^4$, 
$\tilde b_5=b_5\omegref^5$, 
and
$\tilde b_6=b_6\omegref^6$. We obtain:
\begin{equation}
\label{eq:nodimbspwl}
\tilde b_0\tilde\sigma^6+
\tilde b_1\tilde\sigma^5+
\tilde b_2\tilde\sigma^4+
\tilde b_3\tilde\sigma^3+
\tilde b_4\tilde\sigma^2+
\tilde b_5\tilde\sigma +
\tilde b_6 =0,
\end{equation}
with:
\begin{eqnarray}
\tilde b_0\hspace{-2mm}&=&\hspace{-2mm}\gamgas+\chi\gamcos+\twooverbeta,\\
\tilde b_1\hspace{-2mm}&=&\hspace{-2mm}
\epsilon^{-1}\left[
 \tilde\nu\left(\gamgas+\twooverbeta\right)+\tilde \eta\left(1+\chi\gamcos+\twooverbeta\right)
 \right], \\
\tilde b_2\hspace{-2mm}&=&\hspace{-2mm}\left(\gamgas+\chi\gamcos\right)\left(\tomegalfsqd+
\tomegzerosqd+\tcrit\right)-\tomegzerosqd+\twooverbeta\left(\tomegzerosqd+\tcrit\right)+
\epsilon^{-2}\tilde\nu\tilde\eta\left(1+\twooverbeta\right),\\
\tilde b_3\hspace{-2mm}&=&\hspace{-2mm}\epsilon^{-1}
\biggl\{
\left[
\tilde\eta\left(1+\chi\gamcos\right)+\tilde\nu\gamgas
\right]
\hspace{-1mm}
\left(\tomegalfsqd+\tomegzerosqd+\tcrit\right)+
\left(\tilde\nu+\tilde\eta\right)\hspace{-1mm}
\left[
\twooverbeta\left(\tomegzerosqd+\tcrit\right)-\tomegzerosqd
\right]
\biggr\},\qquad\quad\\
\tilde b_4\hspace{-2mm}&=&\hspace{-2mm}\tomegalfsqd
\left[
\left(\gamgas+\chi\gamcos\right)\left(\tomegzerosqd+\tcrit\right)-\tomegzerosqd
\right]
+\epsilon^{-2}\tilde\nu\tilde\eta
\left[
\tomegalfsqd+\crit+\twooverbeta\left(\tomegzerosqd+\tcrit\right)
\right], \\
\tilde b_5\hspace{-2mm}&=&\hspace{-2mm}
\epsilon^{-1}\,\tomegalfsqd\left[
\left(\tomegzerosqd+\tcrit\right)
\left[
\tilde\nu\gamgas+\tilde\eta\left(1+\chi\gamcos\right)
\right]
-\left(\tilde\nu+\tilde\eta\right)\tomegzerosqd
\right]
, \\
\tilde b_6\hspace{-2mm}&=&\hspace{-2mm}\epsilon^{-2}\tilde\nu\tilde\eta\tomegalfsqd\tcrit, 
\end{eqnarray}
As in the previous section we obtain a 6th degree polynomial, and we
anticipate that the solutions will split naturally into two sets, one
in which $\sigma$ is of order the diffusive frequencies and which we
will call the ``diffusive modes,'' and one in which $\sigma$ is of order
the buoyancy frequencies which we will call the ``isothermal buoyancy modes.'' 
We consider each of these sets of solutions in the following subsections.

\subsubsubsection{Diffusive modes}
We expand $\tilde\sigma$ according to:
\begin{equation}
\label{eq:swaveexp}
\tilde\sigma=\epsilon^{-1}\tilde\sigma_{-1}
+\epsilon^0\tilde\sigma_0+\ldots,
\end{equation}
where now $\tsigma_{-1}$ is the lowest-order part of $\tsigma$
and $\tsigma_0$ is the lowest-order correction to $\sigma_{-1}$.
We then substitute the expansion given by equation (\ref{eq:swaveexp})
into equation~(\ref{eq:nodimbspwl}) and collect like terms in $\epsilon$.
The lowest-order terms are of order $\epsilon^{-6}$.
Collecting these and requiring that their sum vanishes
separately we obtain the quadratic:
\begin{equation}
a\sigma_{-1}^2+b\sigma_{-1}+c=0,
\end{equation}
where
\begin{eqnarray}
a&=& \gamgas+\chi\gamcos+\twooverbeta, \\
b&=& \nu\left(\gamgas+\twooverbeta\right) 
         + \eta\left(1+\chi\gamcos+\twooverbeta\right), \\
c&=& \nu\eta\left(1+\twooverbeta\right),
\end{eqnarray}
and where we have once again reverted to the dimensional
form of our solution. The discriminant is
\begin{equation}
b^2-4ac\left[
\nu\left(\gamgas+\twooverbeta\right)-
\eta\left(1+\chi\gamcos+\twooverbeta\right)
\right]^2
+
4\nu\eta\left(\gamgas-1\right)\chi\gamcos,
\end{equation}
and is always positive so that $\sigma_{-1}$ is always real.
Since $b>0$,  the mode is always damped. This result is just
the generalization to the case of arbitrary $\beta$ of the
diffusive mode referred to in section 4.2 of \paperi.

\subsubsubsection{Isothermal buoyancy modes}

For the isothermal buoyancy mode we set
$\tilde\sigma_{-1}=0$ in equation (\ref{eq:swaveexp}) and substitute into
equation~(\ref{eq:nodimbspwl}) again. The terms of
order~$\epsilon^{-2}$ then yield a quadratic equation for~$\sigma_0^2$,
whose solution in dimensional form is
\begin{equation}
\sigma_0^2 = \frac{-b \pm \sqrt{b^2 - 4ac}}{2a},
\label{eq:ibm} 
\end{equation}
where this time
\begin{eqnarray}
a&=&1+\twooverbeta, \\
b&=& \left(\omegalfsqd+\twooverbeta\omegzerosqd\right)
           + \left(1+\twooverbeta\right)\crit, \\
c&=& \omegalfsqd\crit.
\end{eqnarray}
In the high-$\beta$ limit, equation~(\ref{eq:ibm}) reduces
to the results of \paperi.

\subsubsubsection{Comparison with numerical solutions}

Once again we compare our approximate analytical solutions
to the solutions obtained numerically from
equation~(\ref{eq:gendisp}) for a set of parameters chosen to ensure
that the solutions presented fall within the bounds of the
short-parallel-wavelength limit. As before we present
two examples of the unstable (isothermal buoyancy) mode; one
for the case $\beta=1$, and one for the case $\beta=100$. These are shown in
the bottom row of figure \ref{fig:f2}.  For each mode,
the location of the critical wave number $\kcrit$ is marked
with a vertical line. It is again seen that within the limits of
validity of our analytical expressions, ($\left|k_y\right|\gg\kcrit$),
the approximate results and numerical results are in good agreement.
Also as before, we mark the location where $\crit$ passes from negative
to positive values and observe that these solutions again honor the
stability criterion $\crit>0$. Finally we note that for both cases,
the range of unstable wavelengths is the same as for the
long-parallel-wavelength limit while, in the $\beta=1$ case, the
maximum growth rate occurs for $k_{y,{\rm max}}\simeq2.10\perkpc$
where we find $\sigma_{+,+,{\rm max}}\simeq 0.625\permyr$,
whereas for the $\beta=100$ case we find $k_{y,{\rm max}}\simeq 0.112\perkpc$ and
$\sigma_{+,+,{\rm max}}\simeq0.0325\permyr$. Here again we
observe the importance of the magnetic field in determining both
the rate of growth and the range of unstable wave numbers.
Comparing these results to those in section \ref{sec:numcomplong},
we also note that high rates of diffusivity lead to growth rates several
times larger than when these rates are low.

\section{Relation to Shu's (1974) analysis of the Parker instability }
\label{ap:pslimit}

In this appendix we show that the general dispersion relation given by
equation~(\ref{eq:gendisp}) reduces properly to the results obtained
by Parker (1966, 1967) and Shu (1974) when these results are restricted
to the short-wavelength approximation ($kH\gg 1$) assumed throughout
this paper. In the limit considered by these authors, cosmic-ray
diffusivity is taken to be infinite and thermal conduction vanishes,
whence $\eta\rightarrow 0$, and $\nu\rightarrow\infty$. Taking these
limits in equation (\ref{eq:gendisp}) and once again neglecting the
terms involving $g d \ln\rho_0/dz$, our dispersion relation reduces to
\begin{equation}
\label{eq:pslim}
c_0\sigma^6 + c_2\sigma^4+c_4\sigma^2+c_6=0,
\end{equation}
where the coefficients are now
\begin{eqnarray}
c_0 &=& 1, \label{eq:ccofzero}\\
c_2 &=& \omegalfsqd+\left(\gamgas+\twooverbeta\right)\omegssqd, \label{eq:ccoftwo}\\
c_4 &=& \omegssqd
\left[
\left(\gamgas+\twooverbeta\right)\left(\omegzerosqd+\crit\right)-\omegzerosqd +\gamgas\omegalfsqd
\right], \hspace{0.3cm}\mbox{and}\label{eq:ccoffour}\\
c_6 &=& \omegssqd\omegalfsqd\left[\left(\gamgas-1\right)\omegzerosqd+\gamgas\crit\right]. \label{eq:ccofsix}
\end{eqnarray}
We compare this result to the expression obtained by Shu (1974) in the
limit of no rotation and no shear which is given by equation (53) of
that paper. To do this we must take account of the differences in
notation, the fact that Shu's expression is given in dimensionless form,
and most importantly that given the global equilibrium assumed in Shu (1974),
the scale height, $H$, in that paper is consistent with our definition of $H$
given by equation (\ref{eq:defH}). It is also important to note that Shu's 
perturbations include a multiplicative ``envelope'' function which is an
exponentially decreasing function of $z$ above the origin. To account for this
the quantity ``$k$'' as defined in Shu (1974) must be set to zero. With all of
these requirements accounted for we find that Shu's result may be expressed
in our notation according to:
\begin{equation}
\label{eq:shudisp}
s_0\sigma^6+s_2\sigma^4+s_4\sigma^2+s_6=0,
\end{equation}
where the coefficients are
\begin{eqnarray}
s_0 &=& 1, \label{eq:pscofzero} \\
s_2 &=& \omegalfsqd +\left(\gamgas+\twooverbeta\right)\left(\omegssqd-\frac{ik_zp_0}{\rho_0H}\right), \label{eq:pscoftwo}\\
s_4 &=& \omegssqd
\left[
\left(\gamgas+\twooverbeta\right)\left(\omegzerosqd+\crit\right)-\omegzerosqd+
\gamgas\omegalfsqd-\twooverbeta\frac{g}{H}\frac{k_y^2}{k^2}-\gamgas\omegalfsqd\left(\frac{ik_z}{k}\frac{1}{kH}\right)
\right], \hspace{0.3cm}  \mbox{and}   \label{eq:pscoffour}\\
s_6 &=& \omegalfsqd\omegssqd
\left[
\left(\gamgas-1\right)\omegzerosqd+\gamgas\crit -\gamgas\omegalfsqd\left(\frac{ik_z}{k}\frac{1}{kH}\right)
\right]. \label{eq:pscofsix}
\end{eqnarray}
Because $k_z\le k$, the terms in equations (\ref{eq:pscoffour}) and
(\ref{eq:pscofsix}) that involve the factor $1/kH$ must be neglected in the 
short-wavelength limit. Additionally, the term in equation (\ref{eq:pscoftwo})
involving $k_z$ can be seen to be small compared to $\omegssqd$ as follows:
\begin{equation}
\frac{ik_zp_0}{\rho_0H\omegssqd} \sim\frac{ik_z}{k^2H} \sim \frac{ik_z}{k}\frac{1}{kH} \ll 1,
\end{equation}
and similarly, the term in equation (\ref{eq:pscoffour}) involving $k_y^2$
can be seen to be small by comparison to $\gamgas\omegalfsqd$ according to:
\begin{equation}
\frac{(2/\beta)(g/H)(k_y^2/k^2)}{\gamgas\omegalfsqd}=\frac{g}{H}\frac{\rho_0}{p_0}\frac{1}{\gamgas k^2}\sim\frac{1}{k^2H^2}\ll1,
\end{equation}
where the second-to-last relation uses the fact that $g\sim p_0H^{-1}/\rho_0$.
Thus the terms by which the coefficients $s_i$ of Shu (1974) differ from the 
coefficients $c_i$ obtained from our general dispersion relation in the 
Parker/Shu limit and given by equations (\ref{eq:ccofzero})$-$(\ref{eq:ccoffour}),
are just those that must be neglected in the short-wavelength limit to which we
have restricted ourselves in this paper, so that in this limit the results we 
have obtained match those of Shu (1974).

We can informally recover the necessary and sufficient condition for stability
in this limit by applying the method of dominant balance (Bender \& Orszag, 1978)
which we have described in appendix \ref{ap:eigenmodes}, to the dispersion
relation given by equation (\ref{eq:pslim}) with the quantity $\omegssqd$ used
to scale the solutions, and with the quantities $\omegzerosqd$ and $\crit$ taken 
to be small compared to $\omegssqd$. Thus to lowest-order we find that solutions,
$\sigma^2$ to equation (\ref{eq:pslim}) that are of order $\omegssqd$ satisfy:
\begin{equation}
\sigma_{0,{\mathrm s}}^4+c_2\sigma_{0,{\mathrm s}}^2+\gamgas\omegalfsqd\omegssqd=0.
\end{equation}
The discriminant of this expression can be expressed in positive definite form as
\begin{equation}
\left(\omegalfsqd-\gamgas\omegssqd\right)^2+2\twooverbeta\omegssqd\left(\omegalfsqd+
\gamgas\omegssqd\right)+\left(\omegssqd\right)^2\left(\twooverbeta\right)^2 > 0,
\end{equation}
and one can see by inspection of equation (\ref{eq:ccoftwo}) that $c_2$ is
also a positive-definite quantity. Thus the solutions,
\begin{equation}
\sigma^2_{0,{\mathrm s}}=\frac{-c_2\pm\sqrt{c_2^2-4\gamgas\omegalfsqd\omegssqd}}{2},
\end{equation}
are always real and negative, and the modes of order
$\sigma_{0,{\mathrm s}}\sim\omega_{\mathrm s}$ are always stable.

The potentially unstable modes $\sigma_{0,u}$ can be found by seeking solutions
to equation (\ref{eq:pslim}) that satisfy $\sigma^2/\omegssqd\sim\epsilon$ where
$\epsilon\ll 1$, while continuing to assume that $\omegzerosqd$ and $\crit$ are 
also small compared to $\omegssqd$. One finds
\begin{equation}
\sigma_{0,\mathrm{u}}=\pm\sqrt{-\frac{1}{\gamgas}
\left[
\left(\gamgas-1\right)\omegzerosqd+\gamgas\crit
\right]},
\end{equation}
from which we conclude that these modes will be stable if and only if
\begin{equation}
\left(\gamgas-1\right)\omegzerosqd+\gamgas\crit>0,
\end{equation}
which is identical to the necessary and sufficient condition given by Shu (1974)
when the latter is evaluated in the limit $kH\gg1$.

\end{document}